\documentstyle[aps,preprint,eqsecnum,fleqn,floats,epsf]{revtex}

\def \omn#1{\vec{\Omega}_{#1}}
\def \fref#1{fig.\ref{#1}}
\def \cd{{\cal D}}
\newcommand{\k}{{\vec{k}}}
\newcommand{\K}{{\vec{K}}}

\newcommand{\kf}{{k_{\rm F}}}

\def \virg{\;\;,}
\def \point{\;\;.}
\def \up{\uparrow}
\def \down{\downarrow}

\def\chapter#1{\section{#1}}
\def\d{{\rm d}}
\def\e{{\rm e}}
\def\sgn{{\rm sgn}}

\begin{document}
\epsfverbosefalse

\author{H.J. Schulz}

\address{Laboratoire de Physique des Solides\\
Universit\'e Paris--Sud\\
91405 Orsay\\
 France}

\title{Fermi liquids and non--Fermi liquids}

\maketitle

\tableofcontents

\section{Introduction}
Much of the current understanding of solid state physics is based on a
picture of non--interacting electrons. This is clearly true at the
elementary level, but in fact extends to many areas of current research,
examples being the physics of disordered systems or mesoscopic physics. The
most outstanding examples where the non--interacting electron picture fails
are provided by electronic phase transitions like superconductivity or
magnetism. However, more generally one clearly has to understand why the
non--interacting approximation is successful, for example in understanding
the physics of metals where one has a rather dense gas (or liquid) of
electrons which certainly interact via their mutual Coulomb repulsion. A
first answer is provided by Landau's theory of Fermi liquids
\cite{landau_1,landau_2,landau_3} which, starting from the (reasonable but
theoretically unproven) hypothesis of the existence of {\em quasiparticles}
shows that the properties of an interacting system of fermions are
qualitatively similar to that of a non--interacting system. A brief outline
of Landau's theory in its most elementary aspects will be given in the
following section, and a re--interpretation as a fixed point of a
renormalization group will be discussed in sec.\ref{sec:rg}.

A natural question to ask is whether Fermi liquid like behavior is
universal in many--electron systems. The by far best studied example
showing that this is not the case is the one--dimensional interacting
electron gas. Starting with the early work of Mattis and Lieb
\cite{mattis_lieb_bos}, of Bychkov et al. \cite{bychkov_parquet} and
of others is has become quite that in one dimension Landau type
quasiparticles do not exist. The unusual one--dimensional behavior has
now received the name ``Luttinger liquid''. This is still a very active
area of research, and the rest of these notes is devoted to the
discussion of various aspects of the physics of one--dimensional
interacting fermions.

The initial plan for these lectures was considerably wider in scope. It
was in particular considered to include a discussion of the Kondo effect
and its non--Fermi--liquid derivatives, as well as possibly current
theories of strongly correlated fermions in dimension larger than one.
This plan however turned out to be overly ambitious and it was decided
to limit the scope to the current subjects, allowing for sufficiently
detailed lectures. Beyond this limitation on the scope of the lectures,
length restrictions on the lecture notes imposed further cuts. In view
of the fact that there is a considerable and easily accessible
literature on Fermi liquid theory, it seemed best to remain at a rather
elementary level at this point and to retain sufficient space for the
discussion of the more unusual one--dimensional case. It is hoped that
the references, especially in the next and the last section will allow
the interested reader to find sources for further study.

\section{Fermi Liquids}
Landau's Fermi liquid theory is concerned with the properties of a
many--fermion system at low temperatures (much lower than the Fermi
energy) in {\em the normal state}, i.e. in the absence of or at least at
temperatures above any symmetry
breaking phase transition (superconducting, magnetic, or otherwise).
The ideal example for Landau's theory is liquid helium 3, above its
superfluid phase transition, however, the conceptual basis of Landau's
theory is equally applicable to a variety of other systems, in
particular electrons in metals. Quantitative applications are however
more difficult because of a variety of complications, in particular the
absence of translational and rotational invariance and the presence of
electron--phonon interactions, which are not directly taken into account
in Landau's theory. Subsequently, I will first briefly discuss the case
of a noninteracting many--fermion system (the Fermi {\em gas}), and then
turn to Landau's theory of the interacting case (the {\em liquid}),
from a phenomenological point of view and limited to the equilibrium
properties.
A much more detailed and complete exposition of these subjects can be found
in the literature
\cite{pines_nozieres,agd,nozieres,kadanoff_baym,baym_pethick}.

\subsection{The Fermi Gas}
In a noninteracting translationally invariant systems, the
single-particle eigenstates are plane waves
with energy
$\varepsilon_{\k} = \k^2/2m $.
We will always use units
so that $\hbar =1$.
The ground state of an $N$--particle system is the well--known Fermi
sea: all states up to the {\em Fermi wavevector} $\kf$ are filled, all
the other states are empty.
The energy of the last occupied state is the {\em Fermi
energy} $E_{\rm F} = \kf^2/(2m)$.

It is usually convenient to define the Hamiltonian in a way so that the
absolute ground state has a well--defined fixed particle number. This is
achieved simply by including the chemical potential in the definition of
the Hamiltonian, i.e. by writing
\begin{equation}
\label{eq:h0}
H = \sum_{\k} \xi_{\k} n_{\k} \virg
\end{equation}
where $n_\k$ is the usual number operator, $\xi_\k = \varepsilon - \mu$,
and the spin summation is implicit. With this definition
of the Hamiltonian, the elementary excitations of the Fermi gas are
\begin{itemize}
\item addition of a particle at wavevector $\k$ ($\delta n_\k=1$). This
requires
$|k| > \kf$, and thus the energy of this excitation is $\epsilon_\k -
\mu > 0$.
\item destruction of a particle at wavevector $\k$ ($\delta n_\k=-1$),
i.e. creation of a hole. This requires
$|k| < \kf$, and thus the energy is $\mu-\epsilon_\k
 > 0$.
\end{itemize}
These excitations change the total number of particles. Construction of
states at constant particle number is of course straightforward: one
takes one particle from some state $\k$, with $|k|<\kf$, and puts it
into a state $\k'$, with $|\k'|>\kf$. These {\em particle--hole
excitations} are parameterized by the {\em two} quantum numbers
$\k$,$\k'$ and thus form a continuum.
The restriction on the allowed values of $\k,\k'$ insures that all
particle--hole states have positive energy. Higher excited states, i.e.
states with many particles and many holes, are
straightforwardly constructed, the only restriction being imposed by the
Pauli principle.

Thermodynamic quantities are easily obtained and are all determined by
the density of states at the Fermi energy. For example, the specific
heat obeys the well known linear law $C(T) = \gamma T$, with
\begin{equation}
\label{e:gam}
\gamma = \frac{2\pi^2}{3} N(E_{\rm F}) k_{\rm B}^2
\end{equation}
and similarly the (Pauli) spin susceptibility $\chi$ and the
compressibility $\kappa$ are given by
\begin{equation}
\chi  =  2 N(E_{\rm F}) \mu_{\rm B}^2 \virg
\label{e:kap}
\kappa  =  2 N(E_{\rm F}) /\rho^2 \point
\end{equation}
It should be emphasized that eqs. (\ref{e:gam}) to (\ref{e:kap}) are
valid for an arbitrary density of states, in particular in solids where
bandstructure effects can change the electronic dispersion relation
quite drastically. Thus, for noninteracting electrons one expects the
so--called ``Wilson ratio'' $
R_W = \pi^2 k_{\rm B}^2 \chi/(3 \mu_{\rm B}^2 \gamma)$
to be unity, independently of details of the bandstructure. Any
deviation from unity is necessarily an indication of some form of
interaction effect.

\subsection{Landau's theory: phenomenological}
\subsubsection{Basic hypothesis}
Landau's theory is to a large extent based on the idea of a continuous
and one--to--one correspondence
between the eigenstates (ground state {\em and} excited states) of the
noninteracting and the interacting system. For this to be an acceptable
hypothesis it is crucial that the interactions do not lead to any form
of phase transition or symmetry--broken ground state. The standard way to
represent this situation is the so--called ``adiabatic switching on'' of an
interaction, parameterized by a variable $V$. On can easily imagine to
numerically calculate eigenvalues of a Hamiltonian as a function of $V$
(though this may be very difficult in practice). The basic hypothesis of
Landau's theory then is that the ground state energy as well as the energy
of the low--lying states as a function of $V$ do not have any singularities
due to some form of symmetry breaking. Moreover, it should be possible to
label the low--lying states by the same quantum numbers as those of the
noninteracting system, i.e. the number of excited particles and holes and
their respective momenta.

In particular one can consider a state
obtained by adding a particle (with momentum $|\vec{p}| > \kf$) to the
noninteracting ground state:
\begin{equation} \label{eq:p}
|\vec{p},N+1\rangle = a_{\vec{p}}^+ |0,N\rangle \point
\end{equation}
Here $a_{\vec{p}}^+$ is a fermion creation operator for momentum state
$\vec{p}$, and $|0,N\rangle$ is the $N$--particle ground state of the
noninteracting system. Now we add some form of particle--particle
interaction. In a translationally invariant system, interactions conserve
total momentum, and thus even after switching on the interaction the state
still has total momentum $\vec{p}$. However, the interaction of the added
particle with the filled Fermi sea, as well as the interaction of the
particles in the sea amongst themselves, will change the distribution of
particles in $\k$--space, and of course also modify the energy of our
state. The complex formed by the particle added at $\vec{p}$ and the
perturbed distribution of the other particles is called a {\em Landau
quasiparticle}. The Pauli principle implied $|\vec{p}| > \kf$ in the absence
of interactions, and by the continuity hypothesis the same restriction
remains valid in the interacting case. In particular, the value of $\kf$,
which imposes a lower limit on the allowed momentum of the quasiparticle, is
unchanged by the interactions.

Analogous considerations can be performed
for a state obtained by destruction of a particle
(e.g. creation of a hole):
\begin{equation} \label{eq:ph}
|\vec{p},N-1\rangle = a_{\vec{-p}} |0,N\rangle \point
\end{equation}
Note that due to the momentum $-\vec{p}$ the total momentum of this state is
indeed $\vec{p}$. The state obtained after switching  on the interactions
(and which of course still obeys $|\vec{p}|<\kf$)
could be called a quasi--hole, but one generally refers to both
quasi--particles and quasi--holes as quasi--particles.

The quasi--particle concept has a certain number of limitations, mainly due
to the fact that, as will be discussed below, the lifetime of a
quasi--particle is finite. However, for excitations close to $\kf$ one has
$1/\tau \propto (\varepsilon -E_{\rm F})^2$, i.e. the lifetime becomes much
longer
than the inverse excitation energy, and the quasi--particles therefore are
reasonably well defined in the vicinity of the Fermi energy. In practice,
this means that Landau's theory is useful for phenomena at energy scales
much smaller than the Fermi energy, but inapplicable otherwise. In metals,
where $E_{\rm F} \approx 3\ldots 5 eV$, this restriction is not too serious when
one is concerned with thermodynamic or transport properties.
One should also note that the ground state energy itself has important contributions
from states well below $E_{\rm F}$, and therefore is not accessible to Landau's
theory. On the other, the {\em excitations above the ground state} are the
fundamental objects of Landau's theory.

\subsubsection{Equilibrium properties}
To be able to derive physical quantities from the picture of the
low--energy excitations, one needs some information about the
energetics of the quasiparticles and of their interactions. To be
specific, starting from the ground state quasiparticle distribution
\begin{eqnarray}
\nonumber
n_0(k) & = & 1 \; \mbox{if $|\k| < \kf$} \\
\label{eq:n0}
       & = & 0 \; \mbox{if $|\k| > \kf$}
\end{eqnarray}
one
considers changes in quasiparticle occupation
number of the form $n_0(k) \rightarrow n_0(k) + \delta n(k)$, i.e.
$\delta n(k) = 1$ represents an excited quasi--particle, $\delta n(k) =
-1$ an excited quasi--hole (with the notation $k=(\k,\sigma)$, and
$\sigma = \uparrow, \downarrow$ the spin index). The corresponding change
in energy is
\begin{equation}
\label{eq:de}
\delta E = \sum_k \varepsilon_\k^0 \delta n(k)
 + \frac{1}{2\Omega}  \sum_{kk'} f(k,k') \delta n(k) \delta n(k')
\virg
\end{equation}
where the first and second term represent  the energy of a single
quasi--particle and the interaction between quasiparticles,
respectively. To be more precise, we assume that the chemical potential
is included in the Hamiltonian, as in eq.(\ref{eq:h0}). Consequently,
$\varepsilon^0_\k$ vanishes on the Fermi surface, and, given that
we are mainly interested in phenomena in the vicinity of $\kf$, it is
sufficient to retain the lowest order term in an expansion around
$|\k| = \kf$. One thus writes
\begin{equation}
\label{eq:ek0}
\varepsilon_\k^0 = \frac{\kf}{m^*}(|\k| - \kf)
\virg
\end{equation}
thus defining the {\em effective mass} $m^*$. The difference between the
``bare mass $m$ and $m^*$ is due to interaction effects which in
principle can be calculated from a microscopic theory of the system in
question. At the present level $m^*$ will be considered as a
phenomenological parameter.

The energy of a quasi--particle added to the system is easily obtained from
eq.(\ref{eq:de}) by calculating the difference in $\delta E$ between a state
with $\delta n(k) = 1$ and a state with $\delta n(k) = 0$. One finds
\begin{equation} \label{eq:eqp}
\varepsilon_k = \varepsilon^0_\k + \frac{1}{\Omega} \sum_{k'} f(k,k') \delta
n(k') \virg
\end{equation}
i.e. the energy of an added quasi--particle is not just the ``bare''
quasiparticle energy $\varepsilon^0_\k$ but also depends, via the
interaction term, on the presence or not of other quasi--particles.
Given that the non--interacting particles obey Fermi--Dirac statistics, the
quasi--particles do so too, and consequently, the occupation probability of a
quasi--particle state is given by
\begin{equation} \label{eq:nqp}
n(k) = \frac{1}{\e^{\beta \varepsilon_k}+1} \point
\end{equation}
Note that the full quasi--particle energy $\varepsilon_k$ and not the bare
$\varepsilon^0_k$ enters this expression. In principle, $n(k)$ thus has to
be determined selfconsistently from eqs.(\ref{eq:eqp}) and (\ref{eq:nqp}).

For the subsequent calculations, it is convenient to transform the
quasiparticle interaction $f(k,k')$. First, spin symmetric and
antisymmetric $f$--functions are defined via
\begin{eqnarray}
\nonumber
f(\k\uparrow,\vec{k'}\uparrow) & = &
f^s(\k,\k') + f^a(\k,\k') \\
f(\k\uparrow,\vec{k'}\downarrow) & = &
f^s(\k,\k') - f^a(\k,\k')
\end{eqnarray}
Moreover, given the implicit restrictions of the theory, one is only
interested in processes where all involved particles are very close to the
Fermi surface. Under the assumption that the interaction functions are
slowly varying as a function of $\k$, one then can set $|\k| = |\k'| = \kf$.
Because of rotational symmetry, the $f$--functions then can only depend on
the angle between $\k$ and $\k'$, called $\theta$. One can then expand the
$f$--function in a Legendre series as
\begin{equation} \label{eq:fleg}
f^{a,s}(\k,\k') = \sum_{L=0}^\infty f_L^{a,s} P_L(\cos \theta) \virg
\cos \theta = \frac{\k \cdot \k'}{\kf^2} \virg
\end{equation}
where the $P_L$ are the Legendre polynomials. This formula can be inverted
to give
\begin{equation} \label{eq:fll}
f_L^{a,s} = \frac{2L+1}{4\pi} \int {\rm d}^2 \Omega P_L(\cos \theta)
f^{a,s}(\k,\k') \point
\end{equation}
 Finally, one usually puts
these coefficients into dimensionless form by introducing
\begin{equation} \label{eq:FL}
F_L^{a,s} = \frac{\kf m^*}{\pi^2} f_L^{a,s} \point
\end{equation}

We are now in a position to calculate some equilibrium properties. The first
one will be the specific heat at constant volume
\begin{equation} \label{eq:cv}
C_\Omega = \frac{1}{\Omega} \frac{\partial U}{\partial T}
\end{equation}
where $U$ is the internal energy. The temperature--dependent part of $U$
comes from thermally excited quasi--particles, as determined by the
distribution (\ref{eq:nqp}). In principle, in this expression
$\varepsilon_k$ is itself temperature--dependent, because of the temperature
dependent second term in eq.(\ref{eq:eqp}). However, one can easily see
that this term only gives contributions of order $T^2$, and therefore can be
neglected in the low--temperature limit. Consequently, one can indeed
replace $\varepsilon_k$ by $\varepsilon^0_\k$, and then one only has to
replace the bare mass by $m^*$ in the result for a non--interacting system
to obtain
\begin{equation} \label{eq:cv2}
C_\Omega = \frac{m^* \kf}{3} k_{\rm B}^2 T \point
\end{equation}

The spin susceptibility (at $T=0$) is related to the second derivative of the
ground state energy with respect to the (spin) magnetization $M$:
\begin{equation} \label{eq:chi}
\chi = \left[ \Omega \frac{\partial^2 E_0}{\partial M^2} \right]^{-1} \point
\end{equation}
At finite temperature the ground state energy $E_0$ has to be replaced by
the free energy $F(T,N)$.
Spin magnetization is created by increasing the number of $\up$ spin particles
and decreasing the number of $\down$ spins ($M = \mu_{\rm B}(N_\up -N_\down)$),
i.e. by changing the Fermi wavevectors for up and down spins: $\kf
\rightarrow \kf + \delta \kf$ for $\sigma = \up$ and $\kf
\rightarrow \kf - \delta \kf$ for $\sigma = \down$. One has
\begin{eqnarray}
\nonumber
\delta n(\k,\up) & = & 1 \; \mbox{if $\kf < |\k| < \kf + \delta \kf$} \\
\label{eq:dnm}
\delta n(\k,\down) & = & -1 \; \mbox{if $\kf-\delta \kf < |\k| < \kf$}
\end{eqnarray}
and the magnetization is
\begin{equation} \label{eq:M}
M = \Omega \frac{\mu_{\rm B} \kf^2}{\pi^2} \delta \kf \point
\end{equation}
Using the $\delta n$'s from eq.(\ref{eq:dnm}), the contributions from the
first and second term in eq.(\ref{eq:de}) are straightforwardly calculated as
\begin{equation}
\label{eq:dei}
\delta E_1  =  \Omega \frac{\kf^3}{2 \pi^2 m^*} \delta \kf^2 \virg \quad
\delta E_2  =  \Omega \frac{\kf^3}{2 \pi^2 m^*} F_0^a \delta \kf^2  \point
\end{equation}
The coefficient $F_0^a$ appears because the distortion (\ref{eq:dnm}) is
antisymmetric in the spin index and has spherical spatial symmetry ($L=0$).
From eqs.(\ref{eq:chi}) and (\ref{eq:dei})  one obtains the susceptibility as
\begin{equation} \label{eq:chi1}
\chi = \frac{1}{1+F_0^a} \frac{\mu_{\rm B}^2 \kf m^*}{\pi^2} \virg
\end{equation}
Note that here (and contrary to the specific heat) interactions enter not
only via $m^*$ but also explicitly via the coefficient $F_0^a$.
The Wilson ratio then is
\begin{equation} \label{eq:rw1}
R_W = \frac{1}{1 + F_0^a} \point
\end{equation}
Beyond the concrete result obtained here, it should be noted that the
contributions from the two terms in eq.(\ref{eq:de}) are both of order
$\delta \kf^2$, even though formally they seem to be of order $\delta n$
and $\delta n^2$, respectively. This is in fact due to the ``extra
smallness'' coming from the vanishing of $\varepsilon_\k^0$ at the Fermi
surface, whereas $f(k,k')$ remains of course finite. On the other hand,
it should also be clear
that adding terms of order $\delta n^3$ to eq.(\ref{eq:de}) would only lead
to higher order corrections.

Following a similar reasoning, one can obtain a result for the
compressibility of a Fermi liquid. The general definition of $\kappa$ is
\begin{equation} \label{eq:ka}
\kappa = -\frac{1}{\Omega} \frac{\partial \Omega}{\partial P} =
\left[ \Omega \frac{\partial^2 E_0}{\partial \Omega^2} \right]^{-1}
\point
\end{equation}
where at finite temperature one again has to replace the ground state energy
by the free energy. For the calculation it is convenient to replace the
variation of the volume at constant particle number by a corresponding
variation of the particle number at constant volume. One then obtains an
equivalent expression:
\begin{equation} \label{eq:ka2}
\kappa = \left[ \rho^2 \frac{\partial^2 e_0}{\partial \rho^2} \right]^{-1}
\virg
\end{equation}
where $\rho$ is the particle density and $e_0$ is the ground state energy
density. The calculation then is very similar to the one for the spin
susceptibility, only now $\kf$ is changed in the same way for up and down
spins, and consequently $F_0^s$ appear in the final result
\begin{equation} \label{eq:ka3}
\kappa = \frac{m^* \kf}{\pi^2 \rho^2 (1+F^s_0)} \point
\end{equation}

It is also interesting that in a translationally invariant system as we have
considered here, the effective mass is not independent of the interaction
coefficients. This can be seen the following way. Because of Galilean
invariance a system moving at constant velocity $v$ has an extra kinetic
energy $N mv^2/2$. On the other hand, such a system can be represented by
taking some particles from the left side of the Fermi sea to the right side,
i.e. by a $\delta n(k)$ of dipolar symmetry. One can then use this $\delta
n(k)$ to calculate the corresponding change in energy and compare to $N
mv^2/2$. From the comparison one finds
\begin{equation} \label{eq:mm}
\frac{m^*}{m} = 1 + F_1^s/3 \point
\end{equation}

In the present phenomenological framework, the $F$--coefficients are
phenomenological parameters, to be determined from experiment. The standard
example is liquid helium 3, for which one finds
\cite{greywall_landau_1,greywall_landau_2} $m^*/m \approx 3$, $F_0^s
\approx 9$, $F_1^s \approx 5$, $F_0^a \approx -0.7$,
$F_1^a \approx -0.55$. These values indicate rather strong interaction
effects. Note the negative value of $F_0^a$, which corresponds to a strong
enhancement of the spin susceptibility.

\subsection{Conclusion}
The Landau theory of equilibrium properties is of rather limited
quantitative predictive power because
there is no prediction about the actual values of the Landau
parameters. Its principal importance is of conceptual nature: from the
single hypothesis about the existence of quasiparticles it follows that
the low temperature equilibrium properties are very similar to those of
a noninteracting system. In particular, the temperature dependencies are
unaffected, only prefactors are interaction--dependent. Actual
quantitative predictions are obtained when one extends the theory to
nonequilibrium properties, using a Boltzmann
equation.\cite{pines_nozieres} A new phenomenon predicted (and actually
observed in $\rm^3He$ \cite{abel_zero_sound}) is the existence of
collective excitations,
called ``zero sound''. This approach also allows the calculation of the
quasiparticle lifetime and its divergence as the Fermi energy is
approached, as well as the treatment of a number of transport phenomena.

It does not seem generally possible to {\em
derive} Landau's theory starting from some microscopic Hamiltonian, apart
possibly in perturbation theory for small interactions. It is however
possible to formulate the basic hypotheses in terms of microscopic
quantities. In particular, in the microscopic language the existence of
quasiparticles translates into the existence of the {\em quasiparticle
pole} in the one particle Green function:
\begin{equation}\label{eq:qppole}
{\cal G}(\k,i\omega_n) = \frac{z_k}{i\omega_n -%
\varepsilon_\k^0 + i \sgn(\omega_n)%
\tau^{-1}} + \ldots \point
\end{equation}
Here $z_\k$ is the so--called quasiparticle weight, and gives rise to a
jump in the momentum distribution function at $k_F$ of height $z_\k$,
rather than unity in the noninteracting case. Similarly, the Landau
interaction parameters can be related to two--particle vertex parts in
the limit of vanishing momentum transfer. It should
however be emphasized that this line of reasoning provides a
microscopic
interpretation of Landau's picture, rather than proving its correctness.
Similar remarks apply to the calculated diverging quasiparticle lifetime:
this at best shows that Landau's picture is internally consistent.

As already mentioned, the ideal system for the application of Landau's
theory is $\rm ^3He$, which has both short--range interaction and is
isotropic. The application to electrons in metals is more problematic.
First, the interactions are long--ranged (Coulombic). This can however
be accommodated by properly including screening effects. More
difficulties, at least at the quantitative level, arise because metals
are naturally anisotropic. This problem is not of fundamental nature:
even when the Fermi surface is highly anisotropic, an expansion like
eq.(\ref{eq:de}) can still be written down and thus interaction
parameters can be defined. However, a simple Legendre expansion like
eq.(\ref{eq:fleg}) is not in general possible, i.e. the description of
the quasiparticle interaction in terms of a few parameters becomes
impossible. An exception cases with a very nearly spherical Fermi
surface like the alkali metals, where a determination of Landau
parameters can indeed be attempted.\cite{pines_nozieres} It should be
noticed that the difficulties with the Landau description of metals
are not of conceptual nature and in particular do
not invalidate the quasiparticle concept but are rather limitations on
the usefulness of the theory for quantitative purposes.

\section{Renormalization group for interacting fermions}
\label{sec:rg}
In this chapter, we will consider properties of interacting fermions in the
framework of renormalization group theory. This will serve two purposes:
first, the treatment one--dimensional interacting fermions, which will be
considered in detail in the following chapters, gives rise to divergences
which can only be handled by this approach. Results obtained in this way
will be an essential ingredient in the subsequent discussion of ``Luttinger
liquids''. More generally, the renormalization group method will clarify the
status of both Landau's Fermi liquid theory and the Luttinger liquid picture
as renormalization group fixed points, thus establishing a link with a
number of other phenomena in condensed matter physics. We will formulate the
problem in terms of fermion functional integrals, as done by Bourbonnais in
the one--dimensional case \cite{bourbon_couplage} and more recently for two
and three dimensions by Shankar \cite{shankar_short,shankar_revue}. For the
most part, I will closely follow Shankar's notation.

Before considering the interacting fermion problem in detail, let us briefly
recall the general idea behind the renormalization group, as formulated by
Kadanoff and Wilson: one is interested in the statistical mechanics of a
system described by some Hamiltonian $H$. Equilibrium properties then are
determined by the partition function
\begin{equation} \label{eq:z0}
Z = \sum_{\mbox{configurations}} \e^{-\beta H} =
\sum_{\mbox{configurations}} \e^{-S} \virg
\end{equation}
where the second equality defines the {\em action} $S=\beta H$. Typically,
the action
contains degrees of freedom at wavevectors up to some {\em cutoff}
$\Lambda$, which is of the order of the dimensions of the Brillouin
zone. One wishes to obtain an ``effective action'' containing only the
physically most interesting degrees of freedom. In standard phase transition
problems this is the vicinity of {\em the point} $\k=0$, however, for the
fermion problem at hand the {\em surface} $|\k| = \kf$ is relevant, and the
cutoff has to be defined with respect to this surface. In order to
achieve this one proceeds as follows:
\begin{enumerate}
\item Starting from a cutoff--dependent action $S(\Lambda)$ one eliminates
all degrees of freedom between $\Lambda$ and $\Lambda/s$, where $s$ is a
factor larger than unity. This gives rise to a new action $S'(\Lambda' =
\Lambda/s)$.
\item One performs a ``scale change'' $\k \rightarrow s \k$. This brings the
cutoff back to its original value, i.e. one obtains a new action
$S'(\Lambda)$. Because of the degrees of freedom integrated out, coupling
constants (or functions) are changed.
\item One chooses a value of $s$ infinitesimally close to unity: $s =
1+\varepsilon$, and performs the first two steps iteratively. This then
gives rise to differential equations for the couplings, which (in favorable
circumstances) can be integrated until all non--interesting degrees of
freedom have been eliminated.
\end{enumerate}

\subsection{One dimension}
\label{sec:1d}
The one--dimensional case, which has interesting physical applications,
will here be mainly used to clarify the procedure. Let us first consider a
noninteracting problem, e.g. a one--dimensional tight--binding model defined
by
\begin{equation} \label{eq:hti}
H = \sum_k \xi_k a^\dagger_k a^{\phantom{\dagger}}_k \virg \quad \xi_k = -2 t \cos k -\mu \virg
\end{equation}
where $t$ is the nearest--neighbor hopping integral. We will consider the
metallic case, e.g. the chemical potential is somewhere in the middle of the
band. Concentrating on low--energy properties, only states close to the
``Fermi points'' $\pm\kf$ are important, and one can then linearize the
dispersion relation to obtain
\begin{equation} \label{eq:hlin}
H= \sum_{k,r=\pm} v_F(rk-\kf) a^\dagger_{kr} a^{\phantom{\dagger}}_{kr} \virg
\end{equation}
where $v_F= 2 t \sin \kf$ is the Fermi velocity, and the index $r$
differentiates between right-- and left--going particles, e.g. particles
close to $\kf$ and $-\kf$. To simplify subsequent notation, we (i) choose
energy units so that $v_F=1$, (ii) translate $k$-space so that zero energy
is at $k=0$, and (iii) replace the $k$--sum by an integral. Then
\begin{equation} \label{eq:hl2}
H= \sum_{r=\pm} \int_{-\Lambda}^\Lambda \frac{\d k}{2\pi} rk a^\dagger_r(k) a^{\phantom{\dagger}}_r(k)
\point
\end{equation}

For the subsequent renormalization group treatment we have to use a
functional integral formulation of the problem in terms of Grassmann
variables (a detailed explanation of this formalism is given by Negele and
Orland \cite{negele_orland}). The partition function becomes
\begin{equation} \label{eq:zl}
Z(\Lambda) = \int {\cal D}\phi \e^{-S(\Lambda)} \virg
\end{equation}
where ${\cal D}\phi$ indicates functional integration over a set of Grassmann
variables. The action is
\begin{equation} \label{eq:sl1}
S(\Lambda) = \int_0^\beta \d\tau\left\{\sum_{r=\pm} \int_{-\Lambda}^\Lambda \frac{\d
k}{2\pi} \phi_r^*(k,\tau)\partial_\tau\phi^{\phantom{*}}_r(k,\tau) + H(\phi^*,\phi)
\right\} \virg
\end{equation}
where the zero--temperature limit $\beta\rightarrow \infty$ has to be taken,
and $H(\phi^*,\phi)$ indicates the Hamiltonian, with each $a^\dagger$ replaced by
a $\phi^*$, and each $a$ replaced by a $\phi$. Fourier transforming with
respect to the imaginary time variable
\begin{equation} \label{eq:sl2}
\phi_r(k,\tau) = T \sum_{\omega_n} \phi_r(k,\omega_n) \e^{-{\rm
i}\omega_n\tau} \quad (\omega_n = 2\pi(n+ 1/2)T)
\end{equation}
and passing to the limit $T\rightarrow 0$ one obtains the noninteracting action
\begin{equation} \label{eq:sl3}
S_0(\Lambda) = \sum_{r=\pm} \int_{-\infty}^\infty \frac{\d\omega}{2\pi}
\int_{-\Lambda}^\Lambda \frac{\d k}{2\pi} \phi_r^*(k,\omega)[-{\rm
i}\omega+rk] \phi^{\phantom{*}}_r(k,\omega) \point
\end{equation}
We notice that this is diagonal in $k$ and $\omega$ which will greatly
simplify the subsequent treatment. Because of the units chosen, $\omega$ has
units of (length)$^{-1}$ (which we will abbreviate as $L^{-1}$), and then
$\phi_r(k,\omega)$ has units $L^{3/2}$.

We now integrate out degrees of freedom. More precisely, we will integrate
over the strip $ \Lambda/s < |k| < \Lambda$, $-\infty < \omega <\infty$. The
integration over {\em all} $\omega$ keeps the action local in time. One then
has
\begin{equation} \label{eq:zz}
Z(\Lambda) = Z(\Lambda,\Lambda/s) Z(\Lambda/s) \virg
\end{equation}
where $Z(\Lambda,\Lambda/s)$ contains the contributions from the integrated
degrees of freedom, and $Z(\Lambda/s)$, $S_0(\Lambda/s)$ are of the form of
eqs.(\ref{eq:zl}) and (\ref{eq:sl3}). The diagonality of $S_0(\Lambda)$ leads
to the factorized form of eq.(\ref{eq:zz}). Introducing the scale change
\begin{equation} \label{eq:sca}
k' = ks \virg \quad \omega' = \omega s \virg \quad \phi' = \phi s^{-3/2}
\end{equation}
one easily finds that in fact $S_0(\Lambda/s) = S_0(\Lambda)$, i.e. the action
does not change under scale change (or {\em renormalization}), we are at a
{\em fixed point}. One should notice that the scale change of $k$ implies
that $k'$ is quantized in units of $\Delta k' = 2 \pi s/L$, i.e. eliminating
degrees of freedom actually implies that we are considering a shorter
system, with correspondingly less degrees of freedom. This means that even
though the action is unchanged the new $Z(\Lambda)$ is the partition
function of a shorter system. To derive this in detail, one has to take into
account the change in the functional integration measure intervening due to
the scale change on $\phi$.

Before turning to the problem of interactions, it is instructive to consider
a quadratic but diagonal perturbation of the form
\begin{equation} \label{eq:s2}
\delta S_2 = \sum_{r=\pm} \int_{-\infty}^\infty \frac{\d\omega}{2\pi}
\int_{-\Lambda}^\Lambda \frac{\d k}{2\pi} \mu(k,\omega)
\phi_r^*(k,\omega)\phi^{\phantom{*}}_r(k,\omega) \point
\end{equation}
We assume that $\mu(k,\omega)$ can be expanded in a power series
\begin{equation} \label{eq:pow}
\mu(k,\omega) = \mu_{00} + \mu_{10}k+\mu_{01}{\rm i}\omega + \ldots
\end{equation}
Under the scale change (\ref{eq:sca}) one then has
\begin{equation} \label{eq:msca}
\mu_{nm} \rightarrow s^{1-n-m} \mu_{nm} \point
\end{equation}
There now are three cases:
\begin{enumerate}
\item a parameter $m_{nm}$ grows with increasing $s$. Such a parameter is
called {\em relevant}. This is the case for $\mu_{00}$.
\item a parameter remain unchanged ($\mu_{10}$,$\mu_{01}$). Such a
parameter is {\em marginal}.
\item Finally, all other parameter decrease with increasing $s$. These are
called {\em irrelevant}.
\end{enumerate}
Generally, one expects relevant parameter, which grow after elimination of
high--energy degrees of freedom, to strongly modify the physics of the
model. In the present case, the relevant parameter is simply a change in
chemical potential and this doesn't change the physics much (the same is
true for the marginal parameters). One can easily see that another relevant
perturbation is a term coupling right-- and left--going particles of the
form $m(\phi^*_1 \phi^{\phantom{*}}_2 +\phi^*_2 \phi^{\phantom{*}}_1)$. This term in fact does lead to a
basic change: it leads to the appearance of a gap at the Fermi level.

Let us now introduce fermion--fermion interactions. The general form of
the interaction term in the action is
\begin{equation}\label{eq:s4}
 S_I = \int_{k\omega} u(1234) \phi^*(1) \phi^*(2) \phi(3) \phi(4)
\point
\end{equation}
Here $\phi(3)$ is an abbreviation  for $\phi_{r_3}(k_3,\omega_3)$, and
similarly for the other factors. $u$ is an interaction function to be
specified. The integration measure is
\begin{eqnarray}
\nonumber
\int_{k\omega} \ldots &=&\left( \prod_{i=1}^4 \int_{-\infty}^\infty
\frac{\d\omega_i}{2\pi} \int_{-\Lambda}^\Lambda \frac{\d k_i}{2\pi}
\right) \\
\label{eq:mea}
&& \times \delta(k_1+k_2-k_3-k_4)
\delta(\omega_1+\omega_2-\omega_3-\omega_4) \ldots \point
\end{eqnarray}
We now note that the dimension of the integration measure is $L^{-6}$,
and the dimension of the product of fields is $L^6$. This in particular
means that if we perform a series expansion of $u$ in analogy to
eq.(\ref{eq:pow}) the constant term will be $s$--independent, i.e.
marginal, and all other terms are irrelevant. In the following we will
thus only consider the case of a constant ($k$-- and
$\omega$--independent) $u$.

These considerations are actually only the first step in the analysis:
in fact it is quite clear that (unlike in the noninteracting case above)
integrating out degrees of freedom will not in general leave the
remaining action invariant. To investigate this effect, we use a more
precise form of the interaction term:
\begin{eqnarray}
\nonumber
S_I &=& g_1 \int_{k\omega} \sum_{ss'} \phi^*_{s+}(1)  \phi^*_{s'-}(2)
\phi^{\phantom{*}}_{s'+}(3)  \phi^{\phantom{*}}_{s-}(4) \\
\label{eq:g1g2}
&& + g_2 \int_{k\omega} \sum_{ss'} \phi^*_{s+}(1)  \phi^*_{s'-}(2)
\phi^{\phantom{*}}_{s'-}(3)  \phi^{\phantom{*}}_{s+}(4) \point
\end{eqnarray}
Here we have reintroduced spin, and the two coupling constants
$g_1$ and $g_2$ denote, in the original language of eq.(\ref{eq:hlin}),
backward
($(\kf,-\kf)\rightarrow(-\kf,\kf)$) and forward
($(\kf,-\kf)\rightarrow(\kf,-\kf)$) scattering. Note that in the absence
of spin the two processes are actually identical.

\begin{figure}[htb]
\centerline{\epsfxsize 6cm
\epsffile{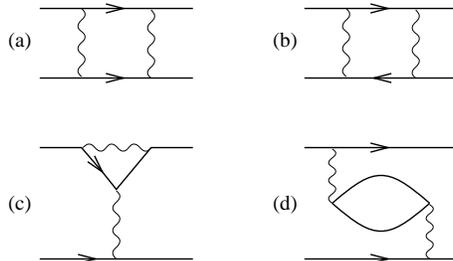}
}
\caption{The diagrams intervening in the renormalization of the coupling
constants $g_1$ and $g_2$. Note that in (b) the direction of one arrow is
reversed with respect to (a), i.e. this is a particle--hole diagram.}
\label{f3:1}
\end{figure}

Now, the Kadanoff--Wilson type mode elimination can be performed via
\begin{equation}\label{eq:esp}
\e^{-S'} = \int {\cal D} \bar{\phi} \e^{-S} \virg
\end{equation}
where $ {\cal D} \bar{\phi}$ denotes integration only over degrees of
freedom in the strip $\Lambda/s < |k| < \Lambda$. Dividing the field
$\phi$ into $\bar{\phi}$ (to be eliminated) and $\phi'$ (to be kept),
one easily sees that the noninteracting action can be written as $S_0 =
S_0(\phi') + S_0(\bar{\phi})$. For the interaction part, things are a
bit more involved: writing
\begin{equation}\label{eq:sI}
S_I = \sum_{i=0}^4 S_{I,i} = S_{I,0} + \bar{S}_I \virg
\end{equation}
where $S_{I,i}$ contains $i$ factors $\bar{\phi}$ we obtain
\begin{equation} \label{eq:spp}
\e^{-S'} = \e^{-S_0(\phi') -S_{I,0}} \int {\cal D} \bar{\phi}
\e^{-S_0(\bar{\phi}) - \bar{S}_I} \point
\end{equation}
Because $\bar{S}_I$ contains up to four factors $\bar{\phi}$, the
integration is not straightforward, and has to be done via a perturbative
expansion, giving
\begin{equation} \label{eq:sbar}
\int {\cal D} \bar{\phi}
\e^{-S_0(\bar{\phi}) - \bar{S}_I} = Z_0(\Lambda,\Lambda/s)
\exp\left[-\sum_{i=1}^\infty \frac{1}{n!} \langle \bar{S}_I^n
\rangle_{\bar{0},con} \right] \virg
\end{equation}
where the notation $\langle \ldots \rangle_{\bar{0},con}$ indicates
averaging over $\bar{\phi}$ and only the connected diagrams are to be
counted. The first order cumulants only give corrections to the energy and
the chemical potential and are thus of minor importance. The important
contributions come from the second order term $\langle S_{I,2}^2
\rangle_{\bar{0},con}$ which after averaging leads to terms of the form
$\phi'^* \phi'^* \phi' \phi'$, i.e. to corrections of the interaction
constants $g_{1,2}$. The calculation is best done diagrammatically, and the
four intervening diagram are shown in fig.\ref{f3:1}.
One can easily see that not all of these diagrams contribute corrections to
$g_1$ or $g_2$. Specifically, one has
\begin{eqnarray}
\nonumber
\delta g_1 & \propto &g_1 g_2 [(a) + (c)] + 2 g_1^2 (d) \\
\label{eq:dg}
\delta g_2 & \propto &(g_1^2+ g_2^2) (a)  +  g_2^2 (b)
\end{eqnarray}
where the factor 2 for diagram $(d)$ comes from the spin summation over the
closed loop. Because the only marginal term is the constant in $u(1234)$,
one can set all external energies and momenta to zero. The integration over
the internal lines in diagram $(a)$ then gives
\begin{equation} \label{eq:inta}
(a) = \int_s \frac{\d k}{2\pi}\int \frac{\d \omega}{2\pi} \frac{1}{{
i}\omega-k} \frac{1}{-{i}\omega-k} = \int_{\Lambda/s}^\Lambda \frac{\d
k}{2\pi} \frac{1}{k} = \frac{1}{2 \pi} \d \ell \virg
\end{equation}
where $s = 1 +\d \ell$, and similarly the particle--hole diagrams $(b)$ to
$(d)$ give a contribution $- d\ell/(2 \pi)$. Performing this procedure
recursively, using at each step the renormalized couplings of the previous
step, one obtains the renormalization group equations
\begin{equation} \label{eq:rg}
\frac{d g_1}{d \ell} = - \frac{1}{\pi} g_1^2(\ell) \virg \quad \frac{d
g_2}{d \ell} = - \frac{1}{2\pi} g_1^2(\ell) \virg
\end{equation}
where $s= \e^\ell$. These equations describe the effective coupling
constants to be used after degrees of freedom between $\Lambda$ and
$\Lambda \e^\ell$ have been integrated out. As initial conditions one of
course uses the bare coupling constants appearing in eq.(\ref{eq:g1g2}).

Equations (\ref{eq:rg}) are easily solved. The
combination $g_1-2g_2$ is $\ell$--independent, and one has further
\begin{equation} \label{eq:gl}
g_1(\ell) = \frac{g_1}{1+g_1 \ell} \point
\end{equation}
There then are two cases:
\begin{enumerate}
\item Initially, $g_1 \ge 0$. One then renormalizes to the {\em
fixed line} $g_1^* = g_1(\ell \rightarrow \infty = 0$, $g_2^* = g_2 -g_1/2$,
i.e. one of the couplings has actually vanished from the problem, but there
is still the free parameter $g_2^*$. A case like this, where perturbative
corrections lead to irrelevancy, is called ``marginally irrelevant''.
\item Initially, $g_1 < 0$. Then $g_1$ diverges at some finite value of
$\ell$. One should however notice that well before the divergence one will
have left the weak--coupling regime where the perturbative calculation
leading to the eq.(\ref{eq:rg}) is valid. One should thus not overinterpret
the divergence and just remember the renormalization towards strong
coupling. This type of behavior is called ``marginally relevant''.
\end{enumerate}
We will discuss the physics of both cases in detail in the next section.

Two remarks are in order here: first, had we done a straightforward
order--by--order perturbative calculation, integrals like eq.(\ref{eq:inta})
would have been logarithmically divergent, both for particle--particle and
particle--hole diagrams. This would have lead to inextricably complicated
problem already at the next order. Secondly, for a spinless problem, the
factor $2$ in the equation for $g_1(\ell)$ is replaced by unity. Moreover,
in this case only the combination $g_1-g_2$ is physically meaningful. This
combination then remains unrenormalized.

\subsection{Two and three dimensions}
\begin{figure}[htb]
\centerline{\epsfxsize 6cm
{\epsffile{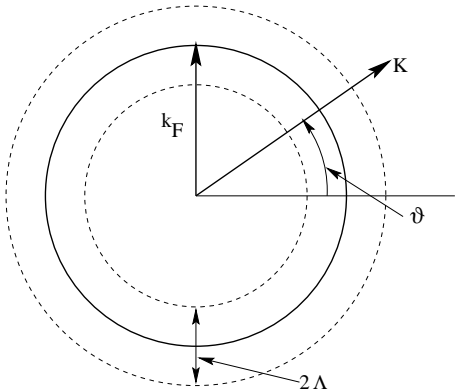}}
}
\caption{Fermi surface geometry in two dimensions.}
\label{f3:2}
\end{figure}

We will now follow a similar logic as above to consider two and more
dimensions. Most arguments will be made for the two--dimensional case, but
the generalization to three dimensions is straightforward. The argument is
again perturbative, and we thus start with free fermions with energy
\begin{equation} \label{eq:xi2d}
\xi_\K = \frac{\K^2}{2m} - \mu = v_F k + O(k^2) \quad (v_F= \kf/m) \point
\end{equation}
We use upper case momenta $\K$ to denote momenta measured from zero, and
lower case to denote momenta measured from the Fermi surface: $k = |\K| -\kf$.
The Fermi surface geometry now is that of a circle as shown in fig.\ref{f3:2}.
One notices in particular that states are now labeled by two quantum numbers
which one can take as radial ($k$) and angular ($\theta$). Note that the
cutoff is applied around the low--energy excitations at $|\K| - \kf$, not
around $\K=0$. The
noninteracting action then takes the form
\begin{equation} \label{eq:s02d}
S_0 = \kf \int_{-\infty}^\infty \frac{\d\omega}{2\pi} \int_0^{2\pi}
\frac{\d\theta}{2\pi} \int_{-\Lambda}^\Lambda\frac{\d k}{2\pi}
[\phi^*(k\theta\omega) (-i \omega - k) \phi(k\theta\omega)]  \point
\end{equation}
One notices that this is just a (continuous) collection of one--dimensional
action functional, parameterized by the variable $\theta$. The prefactor
$\kf$ comes from the two--dimensional integration measure $\d^2 K = (\kf +k)
\d k \d\theta$, where the extra factor $k$ has been neglected because it is
irrelevant, as discussed in the previous section.

The general form of the interaction term is the same as in the
one--dimensional case
\begin{equation} \label{eq:si2d}
 S_I = \int_{\K\omega} u(1234) \phi^*(1) \phi^*(2) \phi(3) \phi(4) \virg
\end{equation}
however, the integration measure is quite different because of
two--dimensional $\K$--space. Performing the integration over $\K_4$ and
$\omega_4$ in the two--dimensional analogue of eq.(\ref{eq:mea}), the
measure becomes
\begin{eqnarray}
\nonumber
\lefteqn{\int_{\K\omega} \ldots =} \\
\label{eq:mea2d}
&& \quad \left( \frac{\kf}{2\pi} \right)^3
\left( \prod_{i=1}^3 \int_{-\infty}^\infty \frac{\d\omega_i}{2\pi} \int_0^{2\pi}
\frac{\d\theta_i}{2\pi} \int_{-\Lambda}^\Lambda\frac{\d k_i}{2\pi} \right)
\Theta(\Lambda-|k_4|) \ldots
\end{eqnarray}
Here $\K_4 = \K_1 + \K_2 - \K_3$.
Now the step function poses a problem because one easily convinces oneself
that even when $\K_{1,2,3}$ are on the Fermi surface, in general $\K_4$ can
be far away from it. This is quite different from the one--dimensional case,
where everything could be (after a trivial transformation) brought back into
the vicinity of $k=0$.

To see the implications of this point, it is convenient to replace the sharp
cutoff in eq.(\ref{eq:mea2d}) by a soft cutoff, imposed by an exponential:
\begin{equation} \label{eq:soft}
\Theta(\Lambda-|k_4|) \rightarrow \exp(-|k_4|/\Lambda) \point
\end{equation}
Introducing now unit vectors $\omn{i}$ in the direction of $\K_i$ via $\K_i
= (\kf+k_i) \omn{i}$ one obtains
\begin{eqnarray}
\nonumber
k_4 & = & |\kf(\omn{1}+\omn{2}-\omn{3}) + k_1 \omn{1}+\ldots| - \kf
\approx \kf (|\vec{\Delta}|-1) \\
\label{eq:k4}
&&\vec{\Delta} = \omn{1}+\omn{2}-\omn{3} \point
\end{eqnarray}
Now, integrating out variables leaves us with $\Lambda \rightarrow
\Lambda/s$ in eq.(\ref{eq:mea2d}) everywhere, including the exponential
cutoff factor for $k_4$. After the scale change (\ref{eq:sca}) the same form
of the action as before is recovered, with
\begin{equation} \label{eq:up}
u'(k_i',\omega_i',\theta_i') = \e^{-(s-1) (\kf/\Lambda) ||\vec{\Delta}|-1|}
u(k/i_s,\omega_i/s,\theta_i) \point
\end{equation}
We notice first that nothing has happened to the angular variable, as
expected because it parameterizes the Fermi surface which is not
affected. Secondly, as in the one--dimensional case, the $k$ and $\omega$
dependence of $u$ is scaled out, i.e. {\em only the values $u(0,0,\theta_i)$
on the Fermi surface are of potential interest (i.e. marginal)}. Thirdly, the
exponential prefactor in eq.(\ref{eq:up}) suppresses couplings for which
$|\vec{\Delta}| \neq 1$. This is the most important difference with the
one--dimensional case.

\begin{figure}[htb]
\centerline{\epsfxsize 6cm
\epsffile{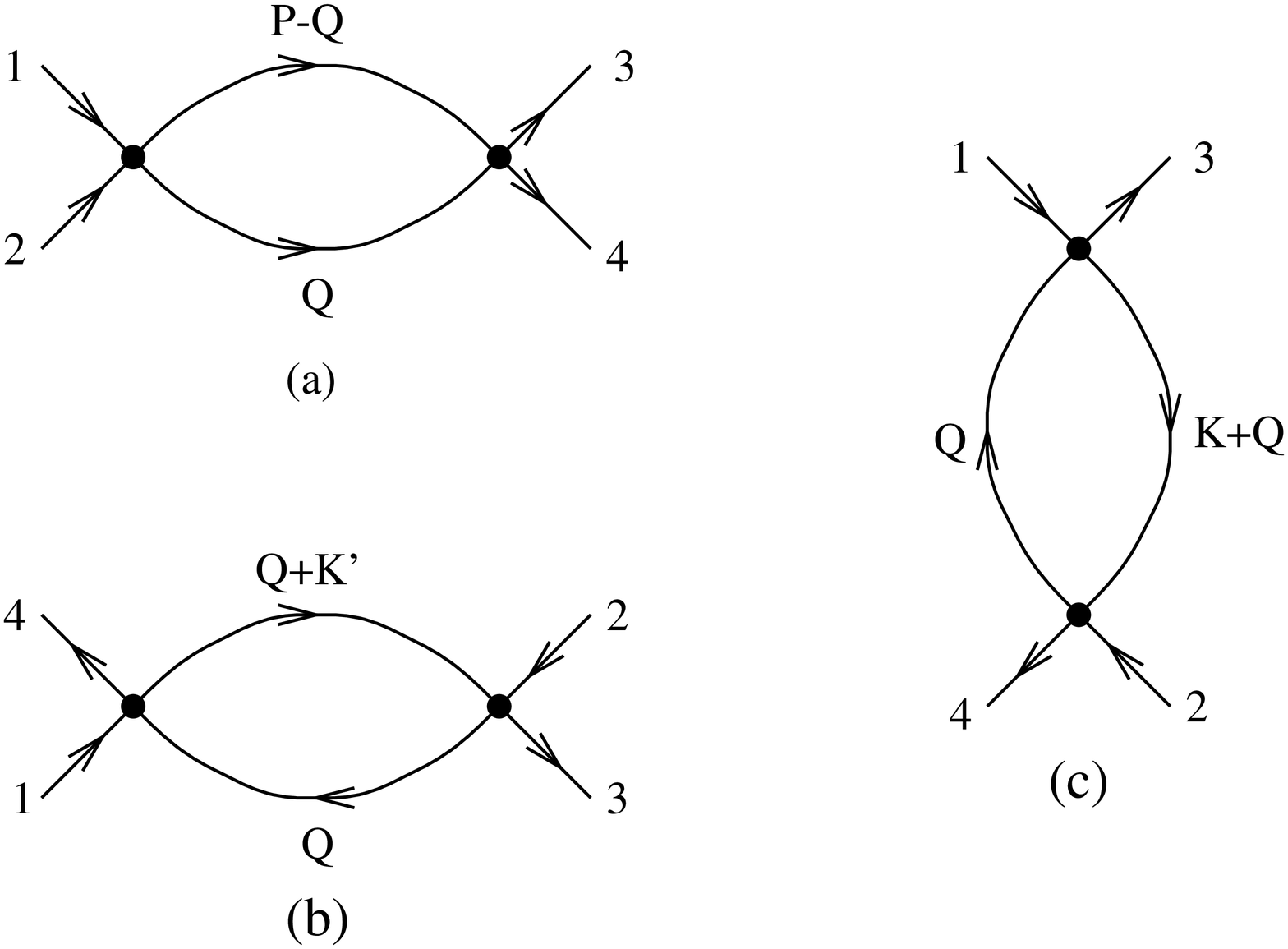}
}
\caption{Second order diagrams renormalizing the coupling function. Here
${\bf P}={\bf K}_1+{\bf K}_2$,${\bf K}_3={\bf K}_1-{\bf K}$, and ${\bf K}_4={\bf K}_1-{\bf K}'$. $\bf Q$ is the loop
integration variable.}
\label{f3:3}
\end{figure}

\begin{figure}[htb]
\centerline{\epsfxsize 6cm
{\epsffile{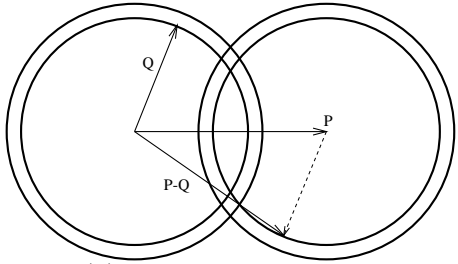}}
}
\caption{Phase space for diagram (a). The rings are the degrees of freedom
to be integrates out between $\Lambda$ and $\Lambda/s$. Note that {\em only}
if ${\bf P}=0$ are ${\bf Q}$ and ${\bf P}- {\bf Q}$ simultaneously in the
area to be integrates, giving a contribution of order $\d\ell$.}
\label{f3:4}
\end{figure}

A first type of solution to $|\vec{\Delta}| = 1$ is
\begin{eqnarray}
\nonumber
\omn{1} = \omn{3} & \Rightarrow & \omn{2} = \omn{4} \virg \mbox{or} \\
\label{eq:pos}
\omn{1} = \omn{4} & \Rightarrow & \omn{2} = \omn{3} \point
\end{eqnarray}
These two cases only differ by an exchange of the two outgoing particles,
and consequently there is a minus sign in the respective matrix
element. Both process depend only on the angle $\theta_{12}$ between
$\omn{1}$ and $\omn{2}$, and we will write
\begin{equation} \label{eq:uf}
u(0,0,\theta_1,\theta_2,\theta_1,\theta_2) =
-u(0,0,\theta_1,\theta_2,\theta_2,\theta_1) = F(\theta_1-\theta_2) \point
\end{equation}
We can now consider the perturbative contributions to the renormalization of
$F$. To lowest nontrivial (second) order the relevant diagrams are those of
Fermi liquid theory and are reproduced in fig.\ref{f3:3}.
Consider diagram (a). To obtain a contribution to the renormalization of
$F$, both $\vec{Q}$ and $\vec{P}-\vec{Q}$ have to lie in the annuli to be
integrates out. As can be seen from fig.\ref{f3:4}, this will give a
contribution of order $\d\ell^2$ and therefore does not contribute to a
renormalization of $F$.
The same is true (if we consider the first case in eq.(\ref{eq:pos})) for
diagram (b). Finally, for diagram (c), because $\K$ is small, the poles of
both intervening Green's functions are on the same side of the real axis,
and here then the frequency integration gives a zero result. For the second
process in eq.(\ref{eq:pos}) the same considerations apply, with the roles
of diagrams (b) and (c) interchanged. The conclusion then is that $F$ is not
renormalized and remains marginal:
\begin{equation} \label{eq:df}
\frac{\d F}{\d \ell} = 0 \point
\end{equation}

The third possibility is to have $\omn{1} = - \omn{2}$,
$\omn{3}=-\omn{4}$. Then the angle between $\omn{1}$ and $\omn{3}$ can be
used to parameterize $u$:
\begin{equation} \label{eq:uv}
u(0,0,\theta_1,-\theta_1,\theta_3,-\theta_3) = V(\theta_1-\theta_3) \point
\end{equation}
In this case $\vec{P}=0$, and therefore in diagram (a) if $\vec{Q}$ is to be
eliminated, so is $-\vec{Q}$. Consequently, one has a contribution of order
$\d\ell$. For the other two diagrams, one finds again negligible
contributions of order $\d \ell^2$. Thus, one obtains
\begin{equation} \label{eq:rgfv}
\frac{\d V(\theta_1-\theta_3)}{\d \ell} = -\frac{1}{8\pi^2} \int_0^{2\pi}
\frac{\d\theta}{2\pi} V(\theta_1-\theta) V(\theta-\theta_3) \point
\end{equation}
This is a renormalization equation for a function, rather than for a
constant, i.e. one here has an example of a ``functional renormalization
group''. Nevertheless, a Fourier transform
\begin{equation} \label{eq:ftv}
V_\lambda = \int_0^{2\pi} \frac{\d\theta}{2\pi} \e^{i\lambda\theta} V(\theta)
\end{equation}
brings this into a more standard form:
\begin{equation} \label{eq:rgv0}
\frac{\d V_\lambda(\ell)}{\d \ell} = - \frac{V_\lambda(\ell)^2}{4\pi} \point
\end{equation}
This has the straightforward solution
\begin{equation} \label{eq:rgv}
V_\lambda(\ell) = \frac{V_\lambda}{1+V_\lambda\ell/(4\pi)} \point
\end{equation}

From eqs.(\ref{eq:df}) and eq.(\ref{eq:rgv}) there are now two
possibilities:
\begin{enumerate}
\item At least one of the $V_\lambda$ is negative. Then one has a divergence
of $V_\lambda(\ell)$ at some finite energy scale. Given that this equation
only receives contributions from BCS--like particle--particle diagrams, the
interpretation of this as a superconducting pairing instability is
straightforward. The index $\lambda$ determines the relative angular
momentum of the particles involved.
\item All $V_\lambda > 0$. Then on has the fixed point $V_\lambda = 0$,
$F(\theta_1-\theta_2)$ arbitrary. What is the underlying physics of this
fixed point? One notices that here one has $\theta_3 = \theta_1$,
$\theta_4=\theta_2$, i.e. the marginal term in the action is
$\phi^*_{\theta_1}\phi^*_{\theta_2}\phi^{\phantom{*}}_{\theta_1}\phi^{\phantom{*}}_{\theta_2}$. In the
operator language, this translates into
\begin{equation} \label{eq:hfp}
H_{int} \approx \int \d\theta_1 \d\theta_2 n_{\theta_1} n_{\theta_2} \point
\end{equation}
We now can recognize this as an operator version of Landau's energy
functional, eq.(\ref{eq:de}). The fixed point theory is thus identified as
{\em Landau's Fermi liquid theory}.
\end{enumerate}
One can notice that the fixed point Hamiltonian, eq.(\ref{eq:hfp}, has a
very large symmetry group: given that $n_k = a^\dagger_ka^{\phantom{\dagger}}_k$, this term (as well
as the kinetic energy term that only contains $n_k$) is invariant under the
U(1) transformation
\begin{equation} \label{eq:uone}
a^{\phantom{\dagger}}_k \rightarrow \e^{i \varphi_k}a^{\phantom{\dagger}}_k
\end{equation}
with arbitrarily $k$--dependent $\varphi_k$, i.e. the symmetry group is
U(1)$^N$, with $N$ the number of points on the Fermi surface. On the other
hand, a standard interaction Hamiltonian has the form
\begin{equation} \label{eq:hig}
H_{int} = \frac{1}{2\Omega} \sum_{kk'q} V(q) a^\dagger_{k+q}
a^\dagger_{k'-q}  a^{\phantom{\dagger}}_{k'} a^{\phantom{\dagger}}_k \virg
\end{equation}
and this is only invariant under a global U(1) transformation.

The generalization of the above to three dimensions is rather
straightforward. In addition to the forward scattering amplitudes $F$,
scattering where there is an angle $\phi_{12;34}$ spanned by the planes
$(\omn{1},\omn{2})$ and $(\omn{3},\omn{4})$ is also marginal. For
$\phi_{12;34}\neq 0$ these processes give rise to a finite
quasiparticle lifetime, however they do
not affect equilibrium properties. The (zero temperature) fixed point
properties thus still only depend on amplitudes for $\phi_{12;34}=0$,
i.e. the Landau $f$--function.

\section{Luttinger liquids}
\label{wcsec}
The Fermi liquid picture described in the preceding two sections is
believed to be relevant for most three--dimensional itinerant electron
systems, ranging from simple metals like sodium to heavy--electron
materials. The most studied example of non--Fermi liquid properties is
that of interacting fermions in one dimension. This subject will be
discussed in the remainder of these lecture notes. We have already
started this discussion in sec.\ref{sec:1d}, where we used a perturbative
renormalization group to find the existence of one marginal coupling,
the combination $g_1-2g_2$. This approach, pioneered by S\'olyom and
collaborators in the early 70's,\cite{solyom_revue_1d} can be extended
to stronger coupling by
going to second or even third order \cite{rezayi_3order} in perturbation
theory. A principal limitation remains however the reliance on
perturbation theory, which excludes the treatment of strong--coupling
problems. An alternative method, which allows one to handle, to a
certain extent, strong--interaction problems as well, is provided by the
bosonization approach, which will be discussed now and which form the
basis of the so--called Luttinger liquid description. It should be
pointed out, however, that frequently entirely analogous results can be
obtained by many--body
techniques.\cite{dzyalo_larkin,evertz_luttinger,dicastro_luttinger}

\subsection{Bosonization for spinless electrons}
\label{sec:bosonspl}
\begin{figure}[htb]
\centerline{\epsfysize 6cm
{\epsffile{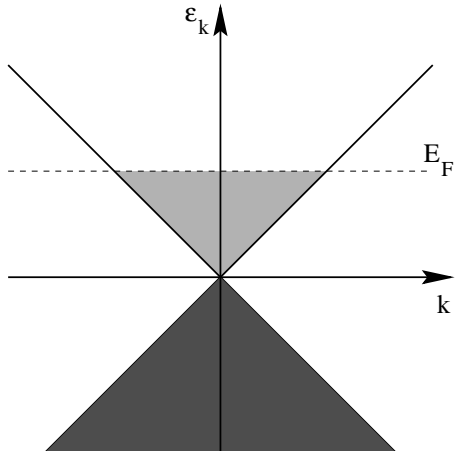}}
}
\caption{Single--particle energy spectrum of the Luttinger model. Occupied
states are shown in grey, the dark grey area represents the states added to
make the model solvable.}
\label{f4:1}
\end{figure}

The bosonization procedure can be formulated precisely, in the
form of operator identities, for fermions with a linear energy--momentum
relation, as discussed in section \ref{sec:1d}. To clarify notation, we
will use $a$--($b$--)operators for right--(left--)moving fermions. The
linearized noninteracting Hamiltonian, eq.(\ref{eq:hlin}) then becomes
\begin{equation}
\label{h0}
H_0 = v_F \sum_{k} \{(k-k_F) a_{k}^\dagger a_{k} +
 (-k-k_F) b_{k}^\dagger b^{\phantom{\dagger}}_{k} \}
\virg
\end{equation}
and the density of states is $N(E_F) = 1/(\pi v_F)$.  In the {\em Luttinger
model},\cite{luttinger_model,mattis_lieb_bos} one generalizes this kinetic
energy by letting the momentum cutoff $\Lambda$ tend to infinity.  There
then are two branches of particles, ``right movers'' and ``left movers'',
both with unconstrained momentum and energy, as shown in
fig.\ref{f4:1}.
At least for weak interaction, this
addition of extra states far from the Fermi energy is not expected to
change the physics much. However, this modification makes the model
exactly solvable even in the presence of nontrivial and
possibly strong interactions.
Moreover, and most importantly, many of the features of this model carry
over even to strongly interacting fermions on a lattice.

We now introduce the Fourier components of
the particle density operator for right and left movers:
\begin{equation}
\rho_+(q) = \sum_k a^\dagger_{k+q} a_k \;\; , \;\;
\rho_-(q) = \sum_k b^\dagger_{k+q} b^{\phantom{\dagger}}_k \;\; ,
\end{equation}
The noninteracting Hamiltonian (and a more general —model including
interactions, see below) can be written in terms of these operators in a
rather simple form and then be solved exactly. This
is based on the following facts:
\begin{enumerate}
\item the density fluctuation operators $\rho_{\pm}$ obey Bose type
commutation relations:
\begin{eqnarray}
\label{eq:b1}
[\rho_+(-q),\rho_+(q')] & = & [\rho_-(q),\rho_-(-q')]
= \delta_{qq'} \frac{qL}{2\pi} \; \; , \quad \\
\label{eq:b2}
[\rho_+(q),\rho_-(q')] & = & 0 \point
\end{eqnarray}
The relation (\ref{eq:b2}) as well as eq.(\ref{eq:b1}) for $q\neq q'$
can be derived by straightforward operator algebra. The slightly
delicate part is eq.(\ref{eq:b1}) for $q=q'$. One easily finds
\begin{equation}\label{eq:b3}
[\rho_+(-q),\rho_+(q)] = \sum_k (\hat{n}_{k-q} - \hat{n}_k) \virg
\end{equation}
where $\hat{n}_k$ is an occupation number {\em operator}. In a usual
system with a finite interval of states between $-\kf$ and $\kf$
occupied, the summation index of one of the $\hat{n}$ operators could be
shifted, giving a zero answer in eq.(\ref{eq:b3}). In the present
situation, with an infinity of states occupied below $\kf$ (see
fig.\ref{f4:2}), this is not so. Consider for example the ground state
\begin{figure}[htb]
\centerline{\epsfxsize 7cm
{\epsffile{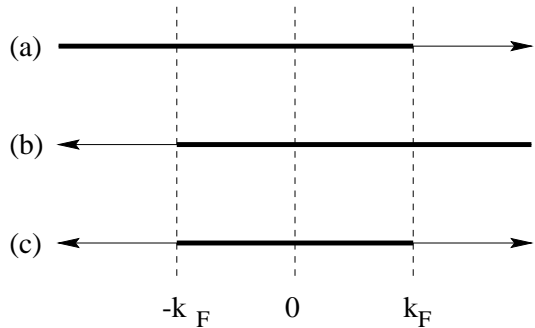}}
}
\caption{Occupied states (heavy line) for right moving fermions (a), left
moving fermions (b), and in a standard model without extra states at
negative energy (c).}
\label{f4:2}
\end{figure}
and $q>0$. Then each term in eq.(\ref{eq:b3}) with $\kf<k<\kf+q$
contributes unity to the sum, all other terms vanish, thus establishing
the result (\ref{eq:b1}). More generally, consider a state with all
levels below a certain value $k_0$ ($<\kf$) occupied, but an arbitrary
number of particle hole pairs excited otherwise. One then has, assuming
again $q>0$,
\begin{eqnarray}\label{eq:b5}
\sum_k (\hat{n}_{k-q} - \hat{n}_k) & = & \left(\sum_{k\geq k_0}
  + \sum_{k<k_0} \right) (\hat{n}_{k-q} - \hat{n}_k) \\
& = & \sum_{k\geq k_0} (\hat{n}_{k-q} - \hat{n}_k) \\
& = & \sum_{k\geq k_0-q}\hat{n}_k-\sum_{k\geq k_0}  \hat{n}_k \\
& = & \sum_{k_0-q\leq k< k_0} \hat{n}_k = \frac{L q}{2\pi}  \point
\end{eqnarray}
The result is independent of $k_0$, and one thus can take the limit
$k_0\rightarrow-\infty$. Together with an entirely parallel argument for
$\rho_-$, this then proves eq.(\ref{eq:b1}).  Moreover, for $q>0$ both
$\rho_+(-q)$ and $\rho_-(q)$ annihilate the noninteracting groundstate. One
can easily recover canonical Bose commutation relations by introducing
normalized operators, e.g. for $q>0$ $b^\dagger_q = \sqrt{2\pi/(qL)}
\rho_+(q)$ would be a canonical creation operator, but we won't use this
type of operators in the following.
\item The noninteracting Hamiltonian obeys a simple commutation relation
with the density operators. For example
\begin{equation}\label{eq:b4}
[H_0,\rho_+(q)] = v_F q \rho_+(q) \virg
\end{equation}
i.e. states created by $\rho_+(q)$ are eigenstates of $H_0$, with energy
$v_F q$.
Consequently, the kinetic part of the Hamiltonian can be re--written as
a term bilinear in boson operators, i.e. quartic in fermion operators:
\begin{equation}
\label{h0b}
H_0 = \frac{2 \pi v_F}{L} \sum_{q>0}
[\rho_+(q) \rho_+(-q) + \rho_-(-q) \rho_-(q)]   \point
\end{equation}
This equivalence may be made more apparent noting that $\rho_+(q)$
creates
particle--hole pairs that all have total momentum $q$. Their energy is
$\varepsilon_{k+q} - \varepsilon_k$, which, because of the linearity of
the spectrum equals $v_F q$, {\em independently of} $k$. Thus, states
created by  $\rho_+(q)$ are linear combinations of individual
electron--hole excitations all with the same energy, and therefore are
also eigenstates of (\ref{h0}).
\item The above point shows that the spectra of the bosonic and
fermionic representations of $H_0$ are the same. To show complete
equivalence, one also has to show that the degeneracies of all the
levels are identical. This can be achieved calculating the partition
function in the two representations and demonstrating that they are
equal. This then shows that the states created by repeated application
of
$\rho_{\pm}$ on the ground state form a complete set of basis states
\cite{haldane_bosonisation,heidenreich_bosonisation}.
\end{enumerate}

We now introduce interactions between the fermions. As long as only
forward scattering of the type
$(k_F;-k_F) \rightarrow
(k_F;-k_F)$ or $(k_F;k_F) \rightarrow  (k_F;k_F)$ is introduced, the
model remains exactly solvable.
The interaction Hamiltonian describing these processes takes the form
\begin{eqnarray}
\nonumber
H_{int} & = & \frac{1}{2L} \sum_{q} \{2
g_2(q) \rho_+(q) \rho_-(-q) \\
\label{hint}
& & \quad + g_4(q) [\rho_+(q) \rho_+(-q) + \rho_-(-q) \rho_-(q)] \}
\point
\end{eqnarray}
Here, $g_2(q)$ and $g_4(q)$ are the Fourier transforms of a real space
interaction potential, and in a realistic case one would of course have
$g_2(q) = g_4(q)$, but it is useful to allow for differences between
$g_2$ and $g_4$.
For Coulomb
interactions one expects $g_2, g_4 > 0$.
In principle, the long--range part of the Coulomb repulsion  leads to a
singular $q$--dependence. Such singularities in the $g_i$ can be handled
rather straightforwardly and can lead to interesting physical effects
as will be discussed below. Here I shall limit myself to nonsingular
$g_2, g_4$.  Electron--phonon
interactions can lead to effectively attractive interactions between
electrons, and therefore in the following I will not make any restrictive
assumptions about the sign of the constants. One should however notice that
a proper treatment of the phonon dynamics and of the resulting retardation
effects requires more care \cite{voit_phonon}.

Putting together (\ref{h0b}) and (\ref{hint}), the complete interacting
Hamiltonian, the {\em Tomonaga--Luttinger model}, then becomes a bilinear form in boson operators that is
easily diagonalized by a Bogolyubov transformation. A first consequence
is the expression for the excitation spectrum
\begin{equation}
\label{omq}
\omega(q) = |q| [(v_F + g_4(q)/(2\pi))^2 - (g_2(q)/(2 \pi))^2]^{1/2}
\point
\end{equation}
The diagonal boson operators are linear combinations of the original
$\rho$ operators, and consequently, these elementary excitations are
collective density oscillations, their energy being determined both by
the kinetic energy term and the interactions.

We note here that in order for the Bogolyubov transformation to be a
well--defined unitary transformation, $g_2(q)$ has to decrease at large
$q$ at least as $|q|^{-1/2}$. on the other hand, the large--$q$
behavior of $g_2$ is unimportant for the low--energy properties of the
model. We therefore in the following will almost always use a
$q$--independent $g_2$ and $g_4$. An approximate and frequently used way to
cure the divergences arising due to this procedure is to keep the
parameter $\alpha$ in subsequent formulae as a finite short--distance
cutoff, of the order of a lattice spacing. One can then also include the
``backward scattering'' $(k_F;-k_F) \rightarrow (-k_F;k_F)$, because for
spinless electron this is just the exchange analogue of forward scattering
and does not constitute a new type of interaction. It is worthwhile
emphasizing here that the solution is valid for arbitrarily strong
interactions, no perturbative expansion is needed!

Up to this point, the construction does not allow for a direct
calculation of correlation functions like the one--particle Green
function or more generally any function involving individual creation or
destruction operators. This type of correlation function becomes
tractable by representing single particle operators in terms of the
boson operators. To this end, we introduce the field operators
\begin{eqnarray}  \label{phi0}
\phi(x) & = & -\frac{i\pi}{L}\sum_{p \ne 0} \frac1{p}
e^{-\alpha |p| /2 -ipx} [\rho_+(p) + \rho_-(p)] - N \frac{\pi x}{L}
\virg \\
\label{Pi0}
\Pi(x) & = & \frac{1}{L}\sum_{p \ne 0}
e^{-\alpha |p| /2 -ipx} [\rho_+(p) - \rho_-(p)] + J /L \\
&& N = N_++N_- \virg \quad J=N_+-N_- \point
\end{eqnarray}
Here $N_\pm$ is the number of particles added to the ground state on the
right-- and left--moving branch. Because addition of a particle changes both
$N$ and $J$, one has the ``selection rule'' $(-1)^N=(-1)^J$, and
$\alpha$ is a cutoff parameter which
(at least in principle, see the discussion above) has to be set to zero in
the end of any calculation. $\phi$ and $\Pi$ then obey canonical boson
commutation relations:
\begin{equation}
[\phi(x),\Pi(y)] = i\delta (x-y) \virg
\end{equation}
and $\phi$ is related to the local particle density via
\begin{equation}
\label{eq:den}
\partial \phi / \partial x = - \pi (\rho(x)-\rho_0) \virg
\end{equation}
where $\rho_0$ is the average particle density in the ground state.
More precisely, in a lattice model this would represent the slowly varying
components ($q\approx 0$) of the density, whereas components with $q\approx
2\kf$ would correspond to crossproducts between $\psi_\pm$.

The expression for the single fermion operators then is
\begin{equation} \label{singlepsi0}
\psi_\pm(x) = \lim_{\alpha \rightarrow 0} \frac1{\sqrt{2\pi
\alpha}} U_\pm \exp \left[
\pm i k_F x \mp i \phi(x) + i \theta(x) \right]
\virg
\end{equation}
where the upper and lower sign refer to right-- and left--moving electrons
respectively, and
\begin{eqnarray}
\nonumber
\theta(x) &= & \pi \int^x \Pi(x') \d x' \\
& = & \frac{i\pi}{L}\sum_{p \ne 0} \frac{1}{p}
e^{-\alpha |p| /2 -ipx} [\rho_+(p) - \rho_-(p)] + J \frac{\pi x}{L} \point
\end{eqnarray}
The $U$--operators decrease the total particle number on one of the
branches by unity and are necessary because the boson fields all conserve
the total particle number.
A detailed derivation of the important eq.(\ref{singlepsi0}) as an
operator identity is given in the literature
\cite{haldane_bosonisation,heidenreich_bosonisation}.
However, a simple plausibility argument can be given:
 creating a particle at site $x$ means
introducing a kink of height $\pi$ in $\phi$, i.e. at points on the left
of $x$ $\phi$ has to be shifted by $\pi$. Displacement operators are
exponentials of momentum operators, and therefore a first guess would be
$\psi(x) \approx \exp (i \pi \int_{-\infty}^x \Pi(x') dx')$.
 However, this operator commutes with
itself, instead of satisfying canonical anticommutation relations.
Anticommutation is achieved by multiplying with an operator, acting at
site $x$, that changes sign each time a particle passes through $x$.
Such an operator is $\exp (\pm i \phi(x))$. The product of these two factors
then produces (\ref{singlepsi0}).

The full Hamiltonian can also be simply expressed in terms of $\phi$ and
$\Pi$. Neglecting the momentum dependence of the $g_i$, one easily finds
\begin{equation}
\label{hbos0}
H = H_0 + H_{int}
= \int dx \left[ \frac{\pi u K}{2} \Pi(x)^2
       + \frac{u}{2\pi K} (\partial_x \phi )^2 \right] \point
\end{equation}
This is obviously just the Hamiltonian of an elastic string, with the
eigenmodes corresponding to the collective density fluctuations of the
fermion liquid. It is important to notice that these collective modes are
the only (low--energy) excited states, and that in particular {\em there are
no well--defined single particle excitations}, nor are there the incoherent
particle--hole pair excitations typical of a Fermi gas. The parameters in
(\ref{hbos0}) are given by
\begin{equation} \label{uk0}
u = [(v_F+g_4/(2\pi))^2 - g_2^2/(2\pi)^2]^{1/2} \virg \;
K = \left[ \frac{2\pi v_F +g_4 - g_2}{2\pi v_F +g_4 + g_2}\right]^{1/2}.
\end{equation}
The energies of the eigenstates are $\omega(q) = u |q|$, in agreement
with eq. (\ref{omq}).

From the continuity equation, the expression (\ref{eq:den}) for the local
particle density and the equation of motion of $\phi$ the (number) current
density is
\begin{equation} \label{eq:jx}
j(x) = uK \Pi(x) \point
\end{equation}
Note in particular that for $g_2= g_4$ one has $u K = v_F$, i.e. the
expression for the current density is interaction--independent. The relation
$u K = v_F$ holds in particular for systems with full (Galilean)
translational invariance. On the other hand, in the continuum limit of
lattice systems this relation is in general not true.

The most remarkable result here is the ``collectivization'' of the
dynamics: there are no quasiparticle--like excitations. In fact there is
a rather simple physical picture explaining this: imagine accelerating
one particle a little bit in one direction. Very soon it will hit its
neighbor and transmit its momentum to it, and the neighbor will in turn
transmit its momentum to a further neighbor, and so on. Quite quickly,
the initial localized motion will have spread coherently through the
whole system. This picture can be formalized noting that in one
dimension the difference between a solid and a fluid is not
well--defined: whereas is higher dimensions solids and fluids are
differentiated by the presence or absence of long--wavelength transverse
modes, no transverse modes can exist in a system with movement along
only one direction. The long--wavelength modes thus can equally well be
considered as the phonons of a one--dimensional
crystal.\cite{haldane_bosons,emery_magog} Note that on the contrary in
dimensions larger than one the neighbors of any given particle can be
pushed aside, giving rise to a backflow that allows the particle to move
trough the system more or less freely.

Rather than discussing the physics of the spinless case in detail, we will
turn now to the more interesting case of fermions with spin.

\subsection{Spin--1/2 fermions}
\label{spinhalfsec}
In the case of spin--1/2 fermions, all the fermion operators acquire an
additional spin index $s$. Following the same logic as above, the kinetic
energy then takes the form
\begin{eqnarray}
\nonumber
H_0 & = & v_F \sum_{k,s} \{(k-k_F) a_{k,s}^\dagger a_{k,s} +
 (-k-k_F) b_{k,s}^\dagger b^{\phantom{\dagger}}_{k,s} \} \\
\label{h0s}
&=& \frac{2 \pi v_F}{L} \sum_{q>0,s}
[\rho_{+,s}(q) \rho_{+,s}(-q) + \rho_{-,s}(-q) \rho_{-,s}(q)]
\virg
\end{eqnarray}
where density operators for spin projections $s=\uparrow,\downarrow$ have
been introduced:
\begin{equation}
\rho_{+,s}(q) = \sum_k a^\dagger_{k+q,s} a_{k,s} \;\; , \;\;
\rho_{-,s}(q) = \sum_k b^\dagger_{k+q,s} b^{\phantom{\dagger}}_{k,s} \;\; \point
\end{equation}
There are now two types of interaction. First, the ``backward scattering''
$(k_F,s;-k_F,t)$ $ \rightarrow $ $ (-k_F,s;k_F,t)$ which for $s \neq t$
cannot
be re--written as an effective forward scattering (contrary to the spinless
case). The corresponding Hamiltonian is
\begin{equation}
H_{int,1}  =  \frac{1}{L} \sum_{kpqst}
g_1 a_{k,s}^\dagger b_{p,t}^\dagger a_{p+2k_F+q,t} b^{\phantom{\dagger}}_{k-2k_F-q,s}
\point
\end{equation}
And, of course, there is also the forward scattering, of a form similar to
the spinless case
\begin{eqnarray}
\label{hint2}
H_{int,2}  & = & \frac{1}{2L} \sum_{qst} \{2
g_2(q) \rho_{+,s}(q) \rho_{-,t}(-q) \\
& & + g_4(q) [\rho_{+,s}(q) \rho_{+,t}(-q) + \rho_{-,s}(-q) \rho_{-,t}(q)]\}
\point
\end{eqnarray}

To go to the bosonic description, one introduces $\phi$ and $\Pi$ fields for
the two spin projections separately, and then transforms to charge and spin
bosons
via $\phi_{\rho,\sigma}= (\phi_\uparrow \pm \phi_\downarrow)/\sqrt2$,
$\Pi_{\rho,\sigma}= (\Pi_\uparrow \pm \Pi_\downarrow)/\sqrt2$.
The operators $\phi_\nu$ and $\Pi_\nu$ obey Bose--like
commutation relations:
$$[\phi_\nu(x),\Pi_{\mu}(y)] = i\delta_{\nu\mu}\delta (x-y) \virg $$
and single fermion operators can be written in a form analogous to
(\ref{singlepsi0}):
\begin{eqnarray}
\nonumber
\lefteqn{\psi_{\pm,s}(x) =} \\
\label{singlepsi}
&& \lim_{\alpha \rightarrow 0} \frac1{\sqrt{2\pi
\alpha}} \exp \left[
{\pm i k_F x}
-i(\pm (\phi_\rho + s \phi_\sigma) + (\theta_\rho+s\theta_\sigma))
/\sqrt{2} \right]
\virg
\end{eqnarray}
where $\theta_\nu(x) = \pi \int^x \Pi_\nu(x') dx'$.

The full Hamiltonian $H = H_0 + H_{int,1} + H_{int,2}$ then takes the form
\begin{equation} \label{hbos}
H = H_\rho + H_\sigma + \frac{2g_1}{(2\pi \alpha )^2}
\int dx \cos (\sqrt8 \phi_\sigma ) \point
\end{equation}
Here $\alpha$ is a short-distance cutoff, and for $\nu = \rho , \sigma $
\begin{equation} \label{hnu}
 H_\nu = \int dx \left[ \frac{\pi u_\nu K_\nu}{2} \Pi_\nu^2
       + \frac{u_\nu}{2\pi K_\nu} (\partial_x \phi_\nu )^2 \right] \virg
\end{equation}
with
\begin{eqnarray}
\nonumber
u_\nu &=& [(v_F+g_{4,\nu}/\pi)^2 - g_\nu^2/(2\pi)^2]^{1/2} \virg \\
\label{uks}
K_\nu & = & \left[ \frac{2\pi v_F +2g_{4,\nu}+ g_\nu}{2\pi v_F +2g_{4,\nu}-
g_\nu}\right]^{1/2} \virg
\end{eqnarray}
and $g_\rho = g_1 -2g_2$, $g_\sigma = g_1$, $g_{4,\rho} = g_4$,
$g_{4,\sigma} = 0$. For a noninteracting system
one thus has $u_\nu = v_F$ (charge and spin velocities equal!) and $K_\nu
=1$. For $g_1=0$,
(\ref{hbos}) describes independent long-wavelength oscillations of the
charge and spin density, with linear dispersion relation $\omega_\nu(k)
= u_\nu |k|$, and the system is conducting. As in the spinless case,
there are no single--particle or single particle--hole pair excited states.
This model (no backscattering), usually called the Tomonaga--Luttinger model,
is the one to which the bosonization method was originally applied
\cite{luttinger_model,mattis_lieb_bos,tomonaga_model}.

For $g_1 \neq 0$ the cosine term has to be treated perturbatively.
We have already obtained the corresponding renormalization group equations
in the previous section (eq.(\ref{eq:rg})). In particular,
 for repulsive interactions ($g_1 >
0$), $g_1$ is renormalized to zero in the long-wavelength limit, and
at the fixed point one has $K_\sigma^* = 1$. The three remaining
parameters in (\ref{hbos}) then completely determine the long-distance
and low--energy properties of the system.

It should be emphasized that (\ref{hbos}) has been derived here for fermions
with linear energy--momentum relation. For more general (e.g.  lattice)
models, there are additional operators arising from band curvature and the
absence of high--energy single--particle states.\cite{haldane_bosonisation}
One can however show that all these effects are, at least for not very
strong interaction, irrelevant in the renormalization group sense, i.e. they
do not affect the low--energy physics. Thus, {\em(\ref{hbos}) is still the
correct effective Hamiltonian for low--energy excitations}.  The lattice
effects however intervene to give rise to ``higher harmonics'' in the
expression for the single--fermion operators, i.e. there are low energy
contributions at wavenumbers $q \approx (2m+1) k_F$ for arbitrary integer
$m$.\cite{haldane_bosons}

The Hamiltonian (\ref{hbos}) also provides an explanation for the physics of
the case of negative $g_1$, where the renormalization group scales to strong
coupling (eq.(\ref{eq:rg})). In fact, if $|g_1|$ is large in
(\ref{hbos}), it is quite clear that the elementary excitations of
$\phi_\sigma$ will be small oscillations around one of the minima of the
$\cos$ term, or possibly soliton--like objects where $\phi_\sigma$ goes from
one of the minima to the other. Both types of excitations have a gap, i.e.
for $g_1 < 0$ one has a {\em gap in the spin excitation spectrum}, whereas
the charge excitations remain massless. This can actually investigated in
more detail in an exactly solvable case.\cite{luther_exact}

\subsubsection{Spin--charge separation}
One of the more spectacular consequences of the Hamiltonian (\ref{hbos})
is the complete separation of the dynamics of the spin and charge
degrees of freedom. For example, in general one has $ u_\sigma \neq
u_\rho$, i.e. the charge and spin oscillations propagate with
different velocities. Only in a noninteracting system or if some
accidental degeneracy occurs does one
have $u_\sigma = u_\rho = v_F$. To make the meaning of this fact more
transparent, let us create an extra particle in the ground state, at
$t=0$ and spatial coordinate $x_0$. The charge and spin densities then
are easily found, using
 $\rho(x) = -(\sqrt2/\pi) \partial \phi_\rho/\partial x$
 (note that $\rho(x)$ is the deviation of the
density from its average value) and
 $\sigma_z(x) = -(\sqrt2/\pi) \partial \phi_\sigma / \partial x$
:
\begin{eqnarray}
\nonumber
\langle 0 | \psi_+(x_0) \rho(x) \psi^\dagger_+(x_0) | 0 \rangle & = &
\delta(x-x_0) \virg \\
\langle 0 | \psi_+(x_0) \sigma_z(x) \psi^\dagger_+(x_0) | 0 \rangle & =
& \delta(x-x_0) \point
\end{eqnarray}
Now, consider the time development of the charge and spin
distributions. The time--dependence of the charge and spin density
operators is easily obtained from  (\ref{hbos}) (using the fixed point
value $g_1 = 0$), and one obtains
\begin{eqnarray}
\nonumber
\langle 0 | \psi_+(x_0) \rho(x,t) \psi^\dagger_+(x_0) | 0 \rangle & = &
\delta(x-x_0-u_\rho t) \virg \\
\langle 0 | \psi_+(x_0) \sigma_z(x,t) \psi^\dagger_+(x_0) | 0 \rangle &
= & \delta(x-x_0-u_\sigma t) \point
\end{eqnarray}
Because in general $u_\sigma \neq u_\rho$, after some time charge and
spin will be localized at completely different points in space, i.e.
{\em charge and spin have separated completely}. An interpretation of
this surprising phenomenon in terms of the Hubbard model will be given
in sec.(\ref{hubsec}). For simplicity we have taken here $K_\rho=1$,
otherwise there would also have been a left--moving part in the
time--dependent density.

Here a linear energy--momentum relation has been
assumed for the electrons, and consequently the shape of the charge
and spin distributions is time--independent. If the energy--momentum
relation has some curvature (as is necessarily the case in lattice
systems) the distributions will widen with time. However this widening
is proportional to $\sqrt{t}$, and therefore much smaller than the
distance
between charge and spin. Thus, the qualitative picture of spin-charge
separation is unchanged.

\subsubsection{Physical properties}
The simple form of the Hamiltonian (\ref{hbos}) at the fixed point
$g_1^*=0$ makes the calculation of physical properties rather
straightforward. The specific heat now is determined both by the charge and
spin modes, and consequently the specific heat coefficient $\gamma$ is given
by
\begin{equation}       \label{gamma}
\gamma/\gamma_0 = \frac12 (v_F/u_\rho + v_F / u_\sigma) \point
\end{equation}
Here $\gamma_0$ is the specific heat coefficient of noninteracting
electrons of Fermi velocity $v_F$.

The spin susceptibility and the compressibility are equally easy to
obtain. Note that in
(\ref{hbos}) the coefficient $u_\sigma / K_\sigma$ determines the energy
necessary to create a nonzero spin polarization, and, as in the spinless
case, $u_\rho
/ K_\rho$ fixes the energy needed to change the particle density. Given
the fixed point value $K_\sigma^*=1$, one finds
\begin{equation}  \label{chi}
\chi / \chi_0 = v_F/u_\sigma \virg \quad \kappa / \kappa_0 = v_F K_\rho
/ u_\rho \virg
\end{equation}
where $\chi_0$ and $\kappa_0$ are the susceptibility and
compressibility of the noninteracting
case.

The quantity $\Pi_\rho(x)$ is proportional to the current density.
As before, the Hamiltonian commutes with the total current, one thus has
\begin{equation}  \label{sig0}
\sigma (\omega) = 2 K_\rho u_\rho \delta (\omega ) +
\sigma_{reg}(\omega) \virg
\end{equation}
i.e. the product $K_\rho u_\rho $ determines the weight of the dc peak in
the conductivity. If the total current commutes with the Hamiltonian
$\sigma_{reg}$ vanishes, however more generally this part of the
conductivity varies as $\omega^3$ at low frequencies.\cite{giamarchi_millis}

\begin{figure}[htb]
\centerline{\epsfysize 6cm
\epsffile{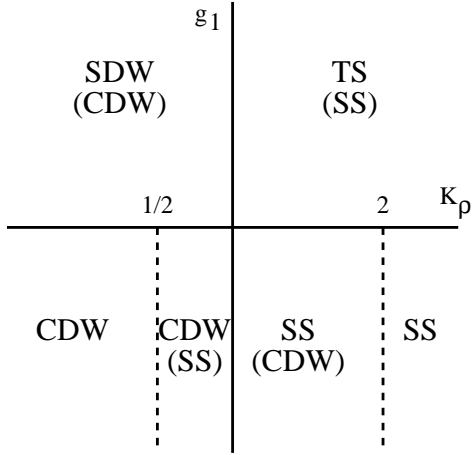}
}
\caption{Phase diagram for interacting spin--$1/2$ fermions.}
\label{f4:4}
\end{figure}

The above properties, linear specific heat, finite spin susceptibility,
and dc conductivity are those of an ordinary Fermi liquid, the
coefficients $u_\rho, u_\sigma$, and $K_\rho$ determining
renormalizations with respect to noninteracting quantities. However, the
present system is {\em not a Fermi liquid}. This is in fact already
obvious from the preceding discussion on charge--spin separation,
and can be made more precise calculating
the single--particle Green function using the
representation (\ref{singlepsi}) of fermion operators.
One then obtains
 the momentum distribution function in the vicinity of $k_F$
\begin{equation}
n_k \approx n_{k_F} -  {\rm const.} \,
 {\rm sign}(k-k_F) |k-k_F|^\delta \;\; \virg
\end{equation}
and  the single-particle density of states (i.e. the
momentum--integrated spectral density):
\begin{equation}
\label{spec}
N(\omega ) \approx |\omega |^\delta \point
\end{equation}
In both cases $\delta=(K_\rho + 1/K_\rho -2)/4$. Note that for any
$K_\rho \neq 1$, i.e. {\em for any nonvanishing interaction}, the
momentum distribution function and the density of
states have power--law singularities at the Fermi level, with a
vanishing single particle density of states at $E_F$. This
behavior is obviously quite different from a standard Fermi liquid
which would have a finite density of states and a step--like singularity
in $n_k$. The absence of a step at $k_F$ in the momentum distribution
function implies the {\em absence of a quasiparticle pole} in the
one--particle
Green function. In fact, a direct calculation of the spectral function
$A(k,\omega)$ \cite{voit_spectral,meden_spectral} shows that the
usual quasiparticle pole is replaced by
a continuum, with a lower threshold at $\min(u_\nu)(k-k_F)$ and branch cut
singularities at $\omega = u_\rho p$ and $\omega = u_\sigma p$:

The coefficient $K_\rho$ also determines the
long-distance decay of all other correlation functions of the system:
Using the representation (\ref{singlepsi}) the charge and spin density
operators at $2k_F$ are
\begin{eqnarray}
\label{ocdw}
O_{CDW}(x) & = & \sum_s \psi_{-,s}^\dagger \psi_{+,s} =
\lim_{\alpha \rightarrow 0}
\frac{e^{2ik_Fx}}{\pi\alpha} e^{-i\sqrt2 \phi_\rho}
\cos [\sqrt2 \phi_\sigma] \virg \\
O_{SDW_x}(x) & = & \sum_{s} \psi_{-,s}^\dagger
\psi_{+,-s}
= \lim_{\alpha \rightarrow 0}
\frac{e^{2ik_Fx}}{\pi\alpha} e^{-i\sqrt2 \phi_\rho}
\cos [\sqrt2 \theta_\sigma] \point
\end{eqnarray}
Similar relations are
also found for other operators. It is important to note
here that all these operators decompose into a product of one factor
depending on the charge variable only by another factor depending only on
the spin field. Using
the Hamiltonian (\ref{hbos}) at the fixed point $g_1^* = 0$ one finds for
example for
the charge and spin correlation functions\footnote{The time- and temperature
dependence is also easily obtained, see
\cite{emery_revue_1d}.}
\begin{eqnarray} \nonumber
\langle n(x) n(0) \rangle & = & K_\rho/(\pi x)^2 + A_1 \cos (2k_Fx)
x^{-1-K_\rho} \ln^{-3/2}(x)  \\
\label{nn}
& & \mbox{} + A_2 \cos (4k_Fx) x^{-4K_\rho} + \ldots \virg \\
\label{ss}
\langle \vec{S}(x) \cdot \vec{S}(0) \rangle & = & 1/(\pi x)^2 + B_1 \cos
(2k_Fx)
x^{-1-K_\rho} \ln^{1/2}(x) + \ldots \virg
\end{eqnarray}
with model dependent constants $A_i,B_i$. The ellipses in (\ref{nn})
and (\ref{ss}) indicate higher harmonics of $\cos (2k_F x)$ which are
present but decay faster than the terms shown here.
Similarly, correlation functions for singlet (SS) and triplet (TS)
superconducting pairing are
\begin{eqnarray}
\nonumber
\langle O^\dagger_{SS}(x) O_{SS}(0) \rangle & = & C x^{-1-1/K_\rho}
\ln^{-3/2}(x) + \ldots \virg \\
\langle O^\dagger_{TS_\alpha}(x) O_{TS_\alpha}(0) \rangle & = & D
x^{-1-1/K_\rho} \ln^{1/2}(x) + \dots \point
\end{eqnarray}
The logarithmic corrections in these functions \cite{finkelstein_logs} have
been studied in detail recently
\cite{voit_logs,giamarchi_logs,affleck_log_corr,singh_logs}.
The corresponding susceptibilities (i.e. the Fourier transforms of the
above correlation functions) behave at low temperatures as
\begin{displaymath}
\chi_{CDW}(T) \approx T^{K_\rho -1 } |\ln(T)|^{-3/2} \virg
\chi_{SDW}(T) \approx T^{K_\rho -1 } |\ln(T)|^{1/2}  \virg
\end{displaymath}
\begin{equation}
\chi_{SS}(T) \approx T^{1/K_\rho -1 } |\ln(T)|^{-3/2} \virg
\chi_{TS}(T) \approx T^{1/K_\rho -1 } |\ln(T)|^{1/2}  \virg
\end{equation}
i.e. for $K_\rho < 1$ (spin or charge) density fluctuations at
$2 k_F$ are enhanced and
diverge at low temperatures, whereas for $K_\rho >1$
pairing fluctuations dominate. The ``phase diagram'', showing in which part
of parameter space which type of correlation diverges for $T\rightarrow 0$
is shown in \fref{f4:4}.

A remarkable fact in all the above results
is that there is only {\em one coefficient}, $K_\rho$, which determines all
the asymptotic power laws,
i.e. there are scaling relations between the exponents of different
correlation functions. These relations would have been impossible
to obtain within a purely perturbative approach.\cite{solyom_revue_1d}
The correlation functions with their power
law decay determine experimentally accessible quantities:
the $2k_F$ and $4k_F$ charge correlations lead to X--ray
scattering intensities $I_{2k_F} \approx T^{K_\rho}$, $I_{4k_F} \approx
T^{4K_\rho-1}$, and similarly the NMR relaxation rate due to $2k_F$ spin
fluctuations varies as $1/T_1 \approx T^{K_\rho}$. Experimental
observations of the NMR relaxation rate \cite{wzietek_nmr} and of
photoemission spectra \cite{dardel_photo} on
the quasi--one--dimensional organic compound $\rm(TMTSF)_2PF_6$ suggest
$K_\rho \approx 0.25$, i.e. rather strong non--Fermi liquid behavior.
It is currently unclear whether this interpretation is compatible with
the transport properties.

We here re--emphasize the two important properties of spin--1/2 interacting
fermions in one dimension: (i) correlation functions
show power--law decay, with
interaction--dependent powers determined by one coefficient, $K_\rho$;
and
(ii) ``spin--charge separation'': spin and charge degrees of freedom
propagate with different velocities. Both these properties invalidate the
Landau quasiparticle concept in one dimension. Rather, the name
``Luttinger liquid'' has been coined to characterize the properties of
one--dimensional interacting fermions.\cite{haldane_bosonisation}

\subsubsection{Long--range interactions}
The above calculations can be straightforwardly generalized to the case of
long--range interactions.\cite{schulz_wigner} Of interest is the case of
unscreened Coulomb
interactions ($V(r) = e^2/r$) which for example is of relevance for the physics
of an isolated quantum wire. The short--distance singularity has to be cut off,
and for example in a wire of diameter $d$ an approximate from would be $V(r) =
e^2/\sqrt{x^2+d^2}$, leading to a Fourier transform $V(q) = g_2(q)
=g_4(q) = 2e^2 K_0(qd)$. The long--range nature of the interaction is
only
of importance for the forward--scattering processes, and these appear only in
the charge part of the Hamiltonian which remains diagonal.
The elementary excitations then are found to be charge oscillations (plasmons),
with energy--momentum relation
\begin{equation}            \label{pla}
\omega_\rho (q) = v_F |q| [(1 + \tilde{g}_1)(1 -
\tilde{g}_1+2\tilde{V}(q))]^{1/2} \end{equation}
where $ \tilde{V}(q) = V(q) / (\pi v_F)$ and
$\tilde{g}_1 = g_1/(2\pi v_F)$. The long--wavelength form,
$\omega_\rho (q) \approx |q^2 \ln q|^{1/2}$, agrees with RPA calculations
\cite{gold_1dplasmon,li_1dplasmon}, however, the effect of $g_1$, which is
a short--range exchange contribution, is usually neglected in those
calculations. The spin modes are still given by  $\omega_\sigma(q) = u_\sigma
|q|$, with $u_\sigma = v_F \sqrt{1-\tilde{g}_1^2}$. Distinct charge--
and spin modes have indeed been observed in Raman scattering from
quantum wires.\cite{goni_gaas1d}

Correlation functions can be similarly calculated and one obtains for example
\begin{eqnarray}
\nonumber
\langle \rho(x) \rho(0) \rangle & = & A_1 \cos(2k_Fx)
\exp(- c_2 \sqrt{\ln x})/x \\
\label{corr}
& & + A_2 \cos(4k_Fx) \exp(- 4 c_2 \sqrt{\ln x}) + ... \;\;,
\end{eqnarray}
where $A_{1,2}$ are interaction dependent constants,
$c_2=\sqrt{(1+\tilde{g}_1)\pi v_F/e^2}$, and
only the most slowly decaying Fourier components are exhibited.
The most interesting point here is the extremely slow decay (much slower
than any power law!) of the $4k_F$ component, showing an incipient charge
density wave at wavevector $4k_F$.
This slow decay should be compared with the power--law decay found for
short--range interactions (eq.(\ref{nn})).
The $4k_F$ oscillation period
is exactly the average interparticle spacing, i.e. the structure is that
expected for a one--dimensional {\em Wigner crystal}. Of course, because of
the one--dimensional nature of the model, there is no true long--range
order, however, the extremely slow decay of the $4k_F$ oscillation would
produce strong quasi--Bragg peaks in a scattering experiment. It is
worthwhile to point out that this $4k_F$ contribution arises even if the
Coulomb interaction is extremely weak and depends only on the long--range
character of the interaction. On the other
hand, any $2k_F$ scattering is considerably weaker, due to the $1/x$
prefactor in (\ref{corr}) which has its origin in the contribution of spin
fluctuations. On the other hand, correlation functions
that involve operators changing the total number of particles (e.g. the
single particle Green function and pairing correlation functions) decay
like $\exp[- cst.(\ln x)^{-3/2}]$,
i.e. {\em faster} than any power law. This in particular means that the
momentum distribution function $n_k$ and all its derivatives are continuous
at $k_F$, and there is only an essential singularity at $k_F$.

It is instructive to notice that the above result (\ref{corr}), obtained
in the
limit of weak Coulomb interactions, can equally be derived assuming
strong long--range repulsion \cite{schulz_wigner} by expanding around
the equilibrium configuration of an equally spaced crystal. We
thus are lead to the rather remarkable conclusion that the
long--distance behavior of
correlation functions is independent of the strength of the Coulomb
repulsion, provided the interaction is truly long--ranged. We note that an
interpretation of Luttinger liquid properties in terms of a one--dimensional
harmonic chain can also be given for short--range interactions
\cite{haldane_bosons,emery_magog,schulz_jeru}.

\subsubsection{Persistent current}
The Luttinger model description can be used straightforwardly to obtain the
current induced in a strictly one--dimensional ring threaded by a magnetic
flux $\Phi$.\cite{loss_ring,fujimoto_hubbard_spinorbit} The argument can in
fact be made very simply: in the one--dimensional geometry, the vector field
can be removed entirely from the Hamiltonian via a gauge transformation,
which then leads to the boundary condition $\psi(L) =
\exp(2\pi i \Phi/\Phi_0) \psi(0)$ for the fermion field operator. Here $L$
is the perimeter of the ring, and $\Phi_0 = hc/e$. For spinless fermions,
this is achieved by replacing
\begin{equation} \label{eq:Pi1}
\Pi(x) \rightarrow \Pi(x) + \frac{2\Phi}{L\Phi_0}
\end{equation}
in the bosonization formula, eq.(\ref{singlepsi0}). The total
$J$--dependent part of the Hamiltonian then becomes{}
\begin{equation}\label{eq:hj}
H_J = \frac{\pi u K}{2 L} (J+ 2\Phi/\Phi_0)^2 \virg
\end{equation}
giving rise to a number current
\begin{equation}\label{eq:jphi}
j = \frac{\Phi_0}{2 \pi} \frac{\partial E}{\partial \Phi} = \frac{uK}{L}
\left(J + \frac{2\Phi}{\Phi_0}\right) \point
\end{equation}
At equilibrium, $J$ is chosen so as to minimize the energy. Given that
at constant total particle number $J$ can only change by two units, one
easily sees that the equilibrium (persistent) current has periodicity
$\Phi_0$, and reaches is maximum value $uK/L$ at $\Phi=\Phi_0/2$,
giving rise to the familiar sawtooth curve for the current as a function
of flux.

For fermions with spin, as long as there is no spin gap ($g_1 >0$), the
above results can be taken over, with the simple replacement $uK
\rightarrow 2 u_\rho K_\rho$, the factor $2$ coming from the spin
degeneracy. Note in particular that the persistent current, an
equilibrium property, is given by the same combination of parameters as
the Drude weight in the conductivity. This is an illustration of Kohn's
result \cite{kohn_drude} relating the Drude weight to the effect of a
magnetic flux through a ring.

In the case of negative $g_1$,
electrons can be transfered from the right to the left--going branch
only by pairs, Consequently, the periodicity of the current and the
ground state energy is doubled to $2\Phi_0$, and the maximum current is
equally doubled. This behavior has actually been found in numerical
calculations.\cite{yu_hubbard_flux,sudbo_cuo}

\section{Transport in a Luttinger liquid}
In the previous section we were concerned with equilibrium properties
and correlation functions, in order to characterize the different phases
possible in a one--dimensional system of interacting fermions. Here, we
will investigate transport, in particular the dc conductivity.
Finite--frequency effects have also been investigated, and the reader is
referred to the literature.\cite{giamarchi_millis,giamarchi_rho}

\subsection{Conductance and conductivity}
To clarify some of the basic notions, let us first consider a Luttinger
model in the presence of a weak space-- and time--dependent external
potential $\varphi$. The interaction of the fermions with $\varphi$ is
described by the extra term
\begin{equation}\label{eq:hext}
H_{ext} = -e \int dx \hat{\rho}(x) \varphi(x,t)
\end{equation}
in the total Hamiltonian. We will assume that the external field is
slowly varying in space, so that in the particle--density operator
$\hat{\rho}$ only products of either two right-- or two left--going
fermion operators appear but no cross terms. Standard linear response
theory tells us that the current induced by the potential is given by
\begin{equation}\label{eq:lr1}
j(x,t) = - \frac{e^2}{\hbar} \int_{-\infty}^t dt' \int dx'
D_{j\rho}(x-x',t-t') \varphi(x',t') \virg
\end{equation}
where the {\em retarded current--density correlation function} is given
by
\begin{eqnarray} \nonumber
D_{j\rho}(x,t) & = & - i \theta(t) \langle[j(x,t),\rho(0,0)]\rangle \\
& = & - \frac{u_\rho K_\rho}{\pi}\theta(t) [\delta'(x-u_\rho t) +
\delta'(x+u_\rho t)]
\point
\label{eq:lr2}
\end{eqnarray}
The second line is the result for spin--$1/2$ electrons. For spinless
fermions one has to make the replacement $u_\rho K_\rho \rightarrow uK/2$.

Let us now first consider the situation where we adiabatically switch on
a potential of frequency $\omega$ and wavenumber $q$ along the whole
length of the system. From eq.(\ref{eq:lr2}) one then straightforwardly
obtains the $q$-- and $\omega$--dependent conductivity as
\begin{equation}\label{eq:sigqo}
\sigma(q,\omega) = \frac{4 e^2}{\hbar} u_\rho K_\rho
\frac{i(\omega+i\delta)}{(\omega+i\delta)^2-u_\rho^2 q^2} \point
\end{equation}
In particular, the real part of the conductivity for constant applied
field is
\begin{equation}\label{eq:sigom}
\sigma(0,\omega) = \frac{2 e^2}{\hbar} u_\rho K_\rho \delta(\omega)
\virg
\end{equation}
in agreement with eq.(\ref{sig0}) (where units with $e^2 = \hbar =1$ were
used).

Applying on the other hand a static field over a finite part of the
sample, one obtains a current $j = 2e^2 K_\rho U/h$,
where $U$ is the applied tension. The conductance thus is
\begin{equation} \label{eq:conduct}
G = \frac{2e^2}{h} K_\rho \virg
\end{equation}
and depends on $K_\rho$ only, not on $u_\rho$. For the noninteracting
case $K_\rho=1$ this is Landauer's well--known
result.\cite{landauer} Note that interactions affect the
value of the conductance. The conductance here is independent of the
length over which the field is applied. Noting that in dimension $d$ the
conductance is related to the $dc$ conductivity via $G = L^{d-2}\sigma$,
a length--independent conductivity implies an infinite conductivity in
one dimension, in agreement with eq.(\ref{eq:lr2}). The fact that
$u_\rho$ does not appear in the expression for $G$ can be understood
noting that applying a static field over a finite (but large) part of
the sample, one is  essentially studying the wavenumber--dependent
conductivity at strictly zero frequency, which from eq.(\ref{eq:lr}) is
given by $\sigma(q,0) =2e^2 K_\rho \delta(q)/\hbar$, indeed
independent
of $u_\rho$. On the other hand, applying a field of finite frequency
over a finite length $\ell$, one can see that one measures the
conductivity $\sigma(q\rightarrow0,\omega)$ if $\omega > u_\rho/\ell$.

\subsection{Edge states in the quantum Hall effect}
\begin{figure}[htb]
\centerline{\epsfxsize 8cm
\epsffile{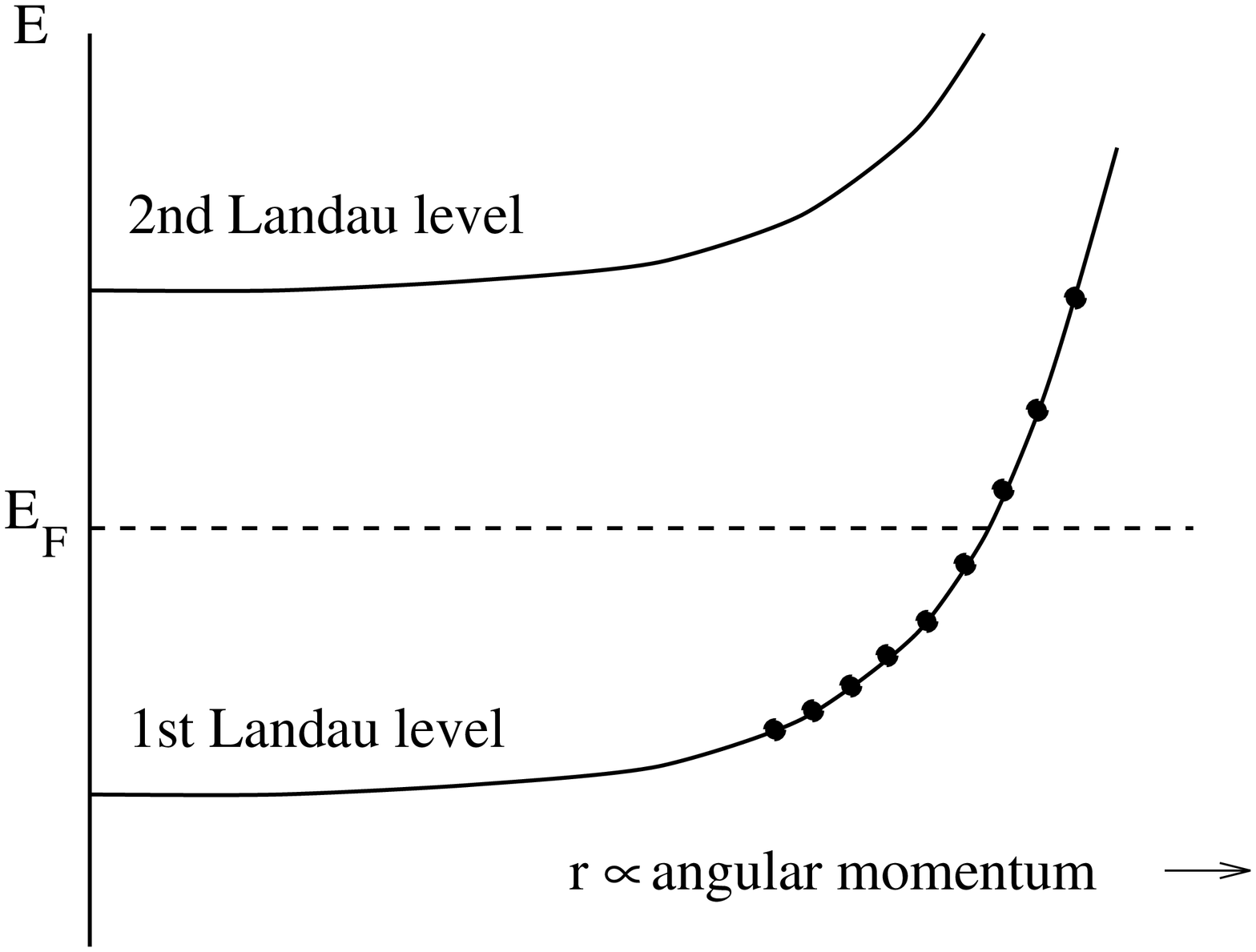}}
\caption[b]{Quantum states in the lowest Landau level in the vicinity of
the edge of a quantum hall device. The spatial variation of the
confining potential is assumed to be slow on the scale of the magnetic
length, so that the energies of the different quantum states are
determined by the local value of the confining potential. For a circular
device the angular momentum of a state increases proportionally to its
distance $r$ from the origin.}
\label{f5:1}
\end{figure}

The Luttinger liquid picture has an interesting application to the
physics of the fractional quantum Hall effect, as discovered and
discussed by Wen.\cite{wen_1,wen_2} To see how this comes about,
consider the states available in the different Landau levels in the
vicinity of the edge of the quantum Hall device,
as shown in fig.\ref{f5:1}.\cite{macdonald_here}
It is clear that low--energy states only exist at the edge (the bulk
quantum Hall state is well-known to be characterized by a finite
excitation gap), and close to the Fermi energy (i.e. the edge) the
states have a linear dispersion relation. This can be made particularly
clear if one assumes a disk--shaped sample: the states have a
wavefunction $$\approx z^k \approx (e^{i\theta})^k \virg$$
with $k$ increasing linearly with radial position. The angular momentum
quantum number $k$ thus plays a role very similar to linear momentum in
the linear geometry we have assumed up to now. One thus can linearize
the the dispersion in fig.\ref{f5:1} and obtains essentially the
spinless model discussed in section \ref{sec:bosonspl}, the only
difference being that here only right--going particles exist. This
difference is the origin of the term {\em chiral} Luttinger liquid (in
fact, the left going branch is to be found on the opposite edge of the
device). Because there are no left--going particles (or at least they
can be thought of as being at a macroscopic distance), there also is no
right--left interaction, and consequently one expects the noninteracting
value $K=1$. Moreover all the left--going components of the fields have
to be projected out, for example one has to replace $\phi \rightarrow
\phi_+=(\phi-\theta)/2$.

However, straightforward adoption of this scheme leads to trouble: from
the preceding subsection we know that $K=1$ leads to a conductance
(which in the present case is the Hall conductance) of $G=e^2/h$, {\em
different} from the well--known
\begin{equation}\label{eq:fraccond}
G = \nu \frac{e^2}{h}
\end{equation}
valid for a fractional quantum Hall state ($\nu = 1/m$ is the filling
factor). To repair this problem one makes the {\em hypothesis} that
instead of eqs.(\ref{eq:den}) and (\ref{eq:jx}) one has
\begin{equation}\label{eq:rhon}
\rho(x) = -\frac{\sqrt{\nu}}{\pi} \frac{\partial \phi_+}{\partial x}
\quad , \quad \quad j(x) = u \sqrt{\nu} \Pi_+(x) \virg
\end{equation}
where the subscripts indicate projection on right--going states. With
these definitions following the calculations of the previous subsection
one now straightforwardly reproduces the correct result,
eq.(\ref{eq:fraccond}). The appearance of the factors $\sqrt{\nu}$ in
eq.(\ref{eq:rhon}) indicates that the objects occupying the states in
fig.\ref{f5:1} are not free electrons but rather strongly affected
by the physics of the bulk of the samples. A more detailed derivation,
starting from a Chern--Simons field theory for the bulk physics, has
also been given by Wen.\cite{wen_2}

Beyond reproducing the correct value of the Hall conductance, the above
hypothesis leads to a number of interesting conclusions. Consider first
the creation operator for a real electron (charge $e$) on the edge.
Following the arguments of section \ref{sec:bosonspl}, because of
eq.(\ref{eq:rhon}), the bosonized version of the electron operator now
must create a jump of $\phi$ of height $\pi/\sqrt{\nu}$, rather than of
height $\pi$. This leads to
\begin{equation}\label{eq:spf}
\psi^{\phantom{\dagger}}_+(x) \approx e^{-i\phi_+(x)/\sqrt{\nu}} \virg
\end{equation}
$x$ being the coordinate along the perimeter of the sample. Now, these
operators obey the relation
\begin{equation}\label{eq:spfcom}
\psi^{\phantom{\dagger}}_+(x') \psi^{\phantom{\dagger}}_+(x) 
= e^{\pm i\pi/\nu} \psi^{\phantom{\dagger}}_+(x) 
\psi^{\phantom{\dagger}}_+(x') \point
\end{equation}
But the real electron is still a fermion, i.e. $\psi^{\phantom{\dagger}}_+$
must obey anticommutation relations. Thus $m=1/\nu$ {\em has to be an odd
integer}. One thus reproduces one of the fundamental facts of the fractional
quantum Hall effect. From eq.(\ref{eq:spf}) one also finds a decay of the
single--electron Green function as
\begin{equation}\label{eq:spfg}
G(x,t) \propto \frac{1}{(x-ut)^{1/\nu}} \point
\end{equation}

Another fundamental property of the quantum Hall state appears when one
considers the fractionally charged elementary excitation of charge
$e\nu$ at the edge. A charge--$e\nu$ object is created by
\begin{equation}\label{eq:spff}
\psi^{\phantom{\dagger}}_{+\nu}(x) \approx e^{-i\sqrt{\nu}\phi_+(x)} \virg
\end{equation}
leading to a slow decay of the corresponding Green function, with
exponent $\nu$, instead of $1/\nu$ in eq.(\ref{eq:spfg}). One now has
the relation
\begin{equation}\label{eq:spfcom2}
\psi^{\phantom{\dagger}}_{+\nu}(x') \psi^{\phantom{\dagger}}_{+\nu}(x) = e^{\pm i\pi\nu} \psi^{\phantom{\dagger}}_{+\nu}(x)
\psi^{\phantom{\dagger}}_{+\nu}(x') \point
\end{equation}
i.e. the fractionally charged objects also obey fractional statistics.

A single hypothesis, the insertion of the factors $\sqrt{\nu}$ in
eq.(\ref{eq:rhon}), thus reproduces two of the fundamental facts about
the fractional quantum Hall effect! In addition one obtains results for
the asymptotics of Green functions.

\subsection{Single barrier}
The infinite conductivity in the ideally pure systems considered up to
here is a natural but hardly realistic result: (almost) any realistic
system will contain some form of inhomogeneity. This in general leads to
a finite conductivity, and in one dimension one can anticipate even more
dramatic effects: in a noninteracting system any form of disorder leads
to localization of the single--particle
eigenstates.\cite{mott_loc,abrahams_loc} How this phenomenon occurs
in interacting systems will be discussed in this and the following
section.

Following Kane and Fisher,\cite{kane_fisher} consider first the case of
a single inhomogeneity in an otherwise perfect one--dimensional system.
The extra term in the Hamiltonian introduced by a localized potential
$v(x)$ is (for spinless fermions)
\begin{equation}\label{eq:hb1}
H_{barrier}  = \int \d x \psi^{\dagger}(x) \psi(x) \point
\end{equation}
Decomposing the product of fermion operators into right-- and
left--going parts, one has
\begin{equation}\label{eq:lr}
\psi^{\dagger} \psi = \psi^{\dagger}_+ \psi^{\phantom{\dagger}}_+ 
+ \psi^{\dagger}_- \psi^{\phantom{\dagger}}_-
+ \psi^{\dagger}_+ \psi^{\phantom{\dagger}}_- + \psi^{\dagger}_- 
\psi^{\phantom{\dagger}}_+ \point
\end{equation}
In the bosonic representation, the first two terms are proportional to
$\partial_x \phi$ (see eq.(\ref{eq:den})), and therefore the
corresponding contribution
in eq.(\ref{eq:hb1}) can in fact be eliminated by a simple unitary
transformation of $\phi$. These terms represent scattering with momentum
transfer $q \ll 2k_F$, i.e. they do not transfer particles between
$k_F$ and $-k_F$ and therefore do not affect the conductance
in any noticeable way. On the other hand, the last two terms in
eq.(\ref{eq:lr}) represent scattering with $|q| \approx 2k_F$, i.e. from
the $+$ to the $-$ branch and vice versa. These terms certainly are
expected to affect the conductance, because they change the direction of
propagation of the particles. The bosonic representation of these terms is
\begin{equation}\label{eq:hb2}
H_{barrier} = \frac{V(2k_F)}{\pi\alpha} \cos 2\phi(0) \virg
\end{equation}
where the potential $V(x)$ is assumed to be centered at $x=0$. For this
reason, only the value of the $\phi$ at $x=0$ intervenes in
eq.(\ref{eq:hb2}).

One now can integrate out all the degrees of freedom away from $x=0$, to
obtain an effective action implying only the time--dependence of
$\phi(0)$. Then a renormalization group equation for $V\equiv V(2k_F)$ can
be found as
\begin{equation} \label{eq:rgv5}
\frac{\d V}{\d \ell} = (1-K) V \virg
\end{equation}
where $E = E_0 e^{-\ell}$, $E_0$ is the original cutoff, and $E$ is the
renormalized cutoff.

From eq.(\ref{eq:rgv5}) it is clear that there are three regimes:
\begin{enumerate}
\item For $K>1$ one has $V(\ell\rightarrow \infty) =0$, i.e. as far as the
low--energy physics is concerned, the system behaves like one without the
barrier. In particular, the low--temperature conductance takes the ``pure''
value $G=e^2 K/h$, with corrections of order $T^{2(K-1)}$. We note that in
this case superconducting fluctuations dominate, and the prefect
transmission through the barrier can be taken as a manifestation of
superconductivity in the one--dimensional system.
\item For the noninteracting case $K=1$, $V$ is invariant, and one thus has
partial transmission and a non--universal conductance depending on $V$.
\item For $K<1$ $V(\ell)$ scales to infinity. Though the perturbative
calculation does not provide any direct way to treat this case, it is
physically clear that the transmission and therefore the conductance should
vanish.
\end{enumerate}
Note that the non--interacting case is marginal, separating the regions of
perfect and zero transmission. These results are very similar to earlier
ones by Luther and Peschel \cite{luther_conductivite_disorder} who treat
disorder in lowest--order perturbation theory.

The case of $K<1$ can be further analyzed considering the case of two finite
Luttinger liquids coupled by a weak tunneling barrier, as would be
appropriate for a strong local potential. The barrier Hamiltonian then
is
\begin{equation} \label{eq:hb3}
H_{barrier} = t[\psi^{\dagger}_1(0) \psi^{\phantom{\dagger}}_2(0)
+ \psi^{\dagger}_2(0) \psi^{\phantom{\dagger}}_1(0)] \approx
\frac{t}{\pi\alpha} \cos 2\theta(0) \point
\end{equation}
Here $\psi^{\phantom{\dagger}}_{1,2}$ are the field operators to the left
and to the right of the barrier. The operators have to satisfy the {\em
fixed boundary condition} $\psi^{\phantom{\dagger}}_i(x=0) =0$, different
from the periodic boundary conditions we have used so far. Noting that the
$\psi^{\phantom{\dagger}}_i$ can be decomposed into left-- and right--going
parts as $\psi^{\phantom{\dagger}}_i = \psi^{\phantom{\dagger}}_{i+} +
\psi^{\phantom{\dagger}}_{i-}$, and using eq.(\ref{singlepsi0} this can be
achieved by imposing the fixed boundary condition $\phi_i(x=0) = \pi/2$ on
the boson field.\cite{eggert_affleck,fabrizio_finiteLL}

One can now proceed in complete analogy to the weak--$V$ case to obtain the
renormalization group equation
\begin{equation} \label{eq:rgt}
\frac{\d t}{\d \ell} = (1-1/K) t \point
\end{equation}
Again, there are three different regimes: (i) for $K>1$ now
$t(\ell\rightarrow\infty) \rightarrow \infty$, i.e. the tunneling amplitude
becomes very big. This can be interpreted as indicating perfect
transmission, e.g. $G=e^2K/h$; (ii) the case $K=1$ remains marginal, leading
to a $t$--dependent conductance; (iii) for $K<1$ $t$ scales to zero, there
thus is no transmission, and $G=0$.

\begin{figure}
\centerline{\epsfxsize 7cm
\epsffile{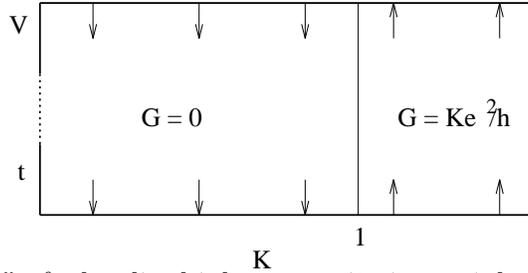}}
\caption[p]{``Phase diagram'' of a localized inhomogeneity in a spinless
Luttinger liquid, characterized by an exponent parameter $K$, according to
\cite{kane_fisher}. The scaling trajectories calculated for weak $V$ or $t$
are indicated by arrows. It is clearly plausible to assume direct scaling
from weak to strong coupling in the whole range of $K$.}
\label{f5:2}
\end{figure}

The results obtained in the two limiting cases of small $V$ (weak
scattering) and of small $t$ (weak tunneling) are clearly compatible:
e.g. for $K<1$ $V$ becomes large, i.e. at sufficiently low energies one
expects essentially a tunneling type behavior, and then from
eq.(\ref{eq:rgt}) the tunneling amplitude actually does scale to zero,
giving zero conductance in the low--energy (or low--temperature) limit. For
$K\ge 1$ a similar compatibility of the two limiting cases is found. The
global behavior can be represented by the ``phase diagram'' in fig.\ref{f5:2}.

For electrons with spin but spin--independent interactions, results are very
similar: the separation between zero and perfect transmission is at $K_\rho
= 1$, with $K_\rho=1$ again the marginal case. In the transmitting region
the conductance is $G=2K_\rho e^2/h$.

These considerations can be generalized to the case of two
barriers.\cite{kane_fisher,furusaki_double_barriere} In particular, assuming
that there are two identical, weakly scattering barrier at $\pm d$, the
effective scattering potential becomes $V_{eff}(q) = 2v(q)
\cos(qd/2)$. Though in general this is non--zero when $V(q)$ is non--zero,
for particular values of $k_F$, so that $\cos(k_F d)=0$ this potential
vanishes, giving rise to perfect transmission even for $K<1$. This {\em
resonant scattering} condition corresponds to an average particle number
between the two barriers of the form $\nu+1/2$, with integer $\nu$, i.e. the
``island'' between the two barriers is in a degenerate state. If
interactions between the electrons in the island are included, one can
recover the physics of the Coulomb
blockade.\cite{kane_fisher,furusaki_double_barriere}

For the chiral Luttinger liquid discussed in the preceding section
backscattering events a priori seem to be excluded because all the
particles are moving in the same direction. In that sense the chiral
Luttinger liquid can be considered as ``perfect''. However, if the
quantum hall device has a constriction that brings the two edges close
to each other, scattering from one edge to the other becomes possible
and is the equivalent of backscattering. Then similar considerations as
made for the single--impurity case are possible \cite{kane_fisher_fqhe},
and in particular the crossover function describing the conductance
through a resonance as a function of temperature has been
obtained.\cite{moon_fqhe}

\subsection{Random potentials and localization}
The discussion of the previous section was concerned with the effect of
a single impurity, weak or strong. Clearly, in that case the effects of
coherent scattering from many impurities which typically give rise to
Anderson localization are absent. We now turn to this more complicated
case which had been studied in fact well before the single impurity
work.\cite{chui_bray,suzumura_loc_scha,apel_loc,giamarchi_loc}

In the absence of electron--electron interactions, localization effects
can be discussed in the framework of a scaling theory.\cite{abrahams_loc}
Under the assumption that at some short length scale one has elastic
scattering of electrons off impurities, this theory leads to the
following $\beta$--function for the variation of the conductance with
linear dimension $L$:
\begin{equation}
\beta (G) = \frac{d \ln (G)}{d \ln (L)} = d-2 - \frac{a}{G} + \dots
\virg
\end{equation}
where $a$ is a constant and $d$ the spatial dimensionality. In
particular in one dimension this leads to a conductance decaying
exponentially with the length of the system, exhibiting clearly the
localized character of the all single--electron states (a fact first
shown by Mott \cite{mott_loc} and studied in great detail since
\cite{erdos_loc1d,efetov_loc}).

In an interacting one--dimensional system (as described by the Luttinger
liquid picture of the previous section) now a number of questions
arise: what is the influence of disorder on the phase diagram obtained
previously? What are the transport properties?
Can one have true superconductivity in one dimension, i.e. infinite
conductivity in a disordered system? To answer these question we
discuss below the generalizations necessary to include disorder in our
previous picture.

We start by the standard term in the Hamiltonian describing the coupling
of a random potential to the electron density
\begin{equation}
H_{imp} = \sum_{R_i} \int dx \; V(x-R_i) \hat{\rho}(x) \virg
\end{equation}
where the $R_i$ are the random positions of impurity atoms, each acting
with a potential $V$ on the electrons. In one dimension one can
distinguish two types of processes:
(i) forward scattering, where the scattered particle remains in the
vicinity of its Fermi point. As in the single--impurity case, this leads
to a term proportional to $\partial \phi_\rho$ and can be absorbed by a
simple redefinition of the $\phi_\rho$ field. The physical effects are
minor, and in particular the dc conductivity remains infinite.
(ii) backward scattering where an electron is scattered from $k_F$ to
$-k_F$ or vice versa. For small impurity density this can be
represented by a complex field
$\xi$ with Gaussian distribution of width $D_\xi = n_i V(q=2k_F)^2$:
\begin{equation}
H_b = \sum_\sigma \int dx \; [\xi(x) \psi^\dagger_{R\sigma}(x)
\psi^{\phantom{\dagger}}_{L\sigma}(x)  + h.c.]
\end{equation}
This term has has dramatic effects and in particular leads to Anderson
localization in the noninteracting case.\cite{abrikosov_loc}

From a perturbative expansion in the disorder one now obtains a set of
coupled renormalization group equations \cite{giamarchi_loc}:
\begin{eqnarray}
\nonumber
\frac{dK_\rho}{d\ell} & = & \lefteqn{- \frac{u_\rho}{2u_\sigma} K_\rho^2
{\cal D}} \\
\nonumber
\frac{dK_\sigma}{d\ell} & = & \lefteqn{-\frac12
({\cal D} + y^2) K^2_\sigma} \\
\nonumber
\frac{dy}{d\ell} & = & \lefteqn{2(1-K_\sigma)y -  {\cal D}} \\
\label{eq:dsca}
\frac{d{\cal D}}{d\ell} & = & \lefteqn{(3-K_\rho-K_\sigma-y){\cal D}}
\virg
\end{eqnarray}
where ${\cal D}= {2D_\xi\alpha}/(\pi u_\sigma^2)
\left({u_\sigma}/{u_\rho}\right)^{K_\rho}$ is the dimensionless
disorder, $y  =
{g_{1\perp}}/({\pi u_\sigma}) $ is the dimensionless backscattering
amplitude, and the $K_\nu$ are defined in eq.(\ref{uks}). These
equations are valid for {\em arbitrary}
$K_\nu$ (the usual strength of bosonization), but to lowest order in
${\cal D}$ and $y$.

\begin{figure}[htbp]
\centerline{\epsfxsize 7cm
\epsffile{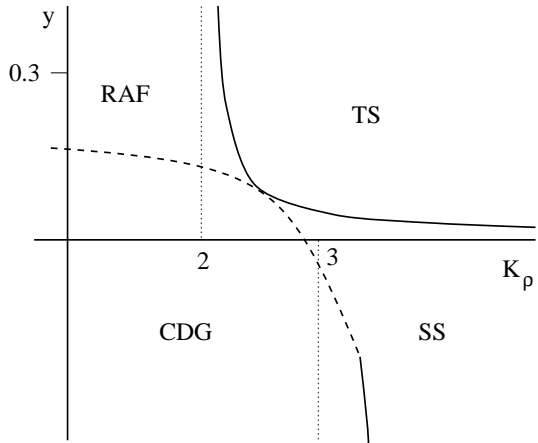}}
\caption[t]{Phase diagram of a Luttinger liquid in the presence of a weak
random potential ($\cd=0.05$). The full lines represent results obtained
directly from the scaling equations (\ref{eq:dsca}), the dashed lines are
qualitative interpolations. The dotted lines are the phase boundaries in the
limit $\cd \rightarrow 0$.}
\label{f5:3}
\end{figure}

As a first application of eqs.(\ref{eq:dsca}) one can determine the
effect of the random potential on the ``phase diagram'', as represented
in fig.\ref{f4:4}. In fact, there are three different regimes:
\begin{enumerate}
\item for $K_\rho > 2$ and  $g_{1\perp}$ sufficiently
positive the fixed point is ${\cal D}^*,y^* =0, K_\rho^* \geq
2$. Because the effective random potential vanishes this is a
delocalized region, characterized as in the pure case by the absence of
a gap in the spin excitations and dominant TS fluctuations.
\item For $K_\rho > 3$ and  $g_{1\perp}$ small or negative one has
${\cal D}^* =0$, $y \rightarrow -\infty $, $K_\rho^* \geq
3$. Again, this is a delocalized region, but now because $y \rightarrow
-\infty$ there is a  spin gap and one has predominant  SS fluctuations.
\item In all other cases one has ${\cal D} \rightarrow \infty,
y \rightarrow -\infty$. This corresponds to a localized regime. For small
$K_\rho$ the bosonized Hamiltonian in this regime is that of a charge
density wave in a weak random potential with small quantum fluctuations
parameterized by $K_\rho$. This region can therefore be identified as a
weakly pinned CDW, also called a ``charge density glass'' (CDG).  The
transition from the CDG to the SS region then can be seen as depinning of
the CDW by quantum fluctuations.
\end{enumerate}
One should notice that the CDG is a nonmagnetic spin singlet, representing
approximately a situation where localized single--particle states are doubly
occupied. Though this is acceptable for attractive or possibly weakly
repulsive interactions, for strong short--range repulsion single occupancy
of localized states seems to be more likely.  One then has a spin in each
localized state, giving probably rise to a localized antiferromagnet with
random exchange (RAF). A detailed theory of the relative stability of the
two states is currently missing and would certainly at least require
higher-order perturbative treatment.  The boundaries of the different
regimes can be determined in many cases from eqs.(\ref{eq:dsca}), and the
resulting phase diagram is shown in fig.\ref{f5:3}.

The localization length for small disorder can be obtained from standard
scaling arguments:
suppose that a system with some fixed disorder $\cd_0$ has a
localization length $\xi_0$. Then in the general case one has
$\xi ( {\cal D}) = \xi_0 e^{\ell (\cd_0,\cd)}$, where $\ell(\cd_0,\cd)$
is the time it takes for the ``bare'' disorder $\cd$ to scale up to
$\cd_0$. From this reasoning one finds, for the case without a spin gap
($g_1>0$) and weak disorder
\begin{equation}
\label{eq:xi1}
\xi (\cd )  \propto  (1/\cd )^{1/(2-K_\rho)}\point
\end{equation}
Note that for $K_\rho>1$, i.e. superconducting fluctuations
predominating in the pure case $\xi$ is greater than the mean free path
$\lambda \propto 1/\cd$, i.e. there is a kind of diffusive regime,
contrary to the noninteracting case. On the other hand, for $K_\rho<1$
one has $\xi<\lambda$. In the vicinity of the TS--RAF boundary one has
\begin{equation}
\label{eq:xi2}
\xi (\cd )  \propto  \exp \left( \frac{K_\rho -2}{\cd - y (K_\rho -2)}
\right) \point
\end{equation}

The analogous results for the case with a spin gap ($g_1<0$) are
\begin{eqnarray}
\label{eq:xi3}
\xi (\cd ) & \propto & (1/\cd )^{1/(3-K_\rho)} \\
\label{eq:xi4}
\xi (\cd ) & \propto & \exp \left( \frac{2 \pi}{\sqrt{9 \cd - (K_\rho
-3)^2}} \right)
\end{eqnarray}
There are two points to be noted about this result: (i) for $K_\rho=0$ one
has $\xi \propto \cd^{-1/3}$, which is the same result as that found for the
pinning length of a classical CDW.\cite{fukuyama_pinning} (ii) the results
(\ref{eq:xi1}), (\ref{eq:xi2}) and (\ref{eq:xi3}), eq.(\ref{eq:xi4}) are
qualitatively different, both in the vicinity of the phase boundaries and in
the localized states. The transitions are thus in different universality
classes, and this strongly supports the idea that the localized phases
reached through the transition are themselves different (RAF or CDG).

The temperature dependence of the dc conductivity can be obtained noting
that at finite temperature there are no coherent effects on length scales
larger than $v_F/T$. One therefore stops renormalization at $e^{\ell^*} =
v_F/(\alpha T)$. As long as $\cd$ remains weak one can still use the Born
approximation to obtain
\begin{equation}
\sigma(\cd)  = \sigma_0 \frac{e^\ell \cd}{\cd(\ell)} \virg
\end{equation}
where $\sigma_0 = e^2v_F^2/2\pi\hbar D_\xi$ is the lowest order
conductivity.  In the delocalized phases one then finds a conductivity
diverging as $ \sigma(T) \sim T^{-1-\gamma} $ where $\gamma = K_\rho^*-2$ in
the TS case and $\gamma = K_\rho^*-3$ in the SS case. On the phase
boundaries one has universally $\sigma \sim 1/T$.  In the localized region
$\cd$ diverges at low temperatures, and a perturbative calculation thus
becomes meaningless. However, the conducting--localized crossover can still
be studied at not too low temperatures.\cite{giamarchi_loc} In particular,
the high--temperature conductivity is found to vary as $\sigma \sim
T^{1-K_\rho}$. This is the perturbative result first found by Luther and
Peschel \cite{luther_conductivite_disorder} and also reproduced by the
single--impurity calculations.\cite{kane_fisher} The high--temperature
behavior thus can be understood in terms of scattering off the individual
impurities. On the other hand, at lower temperatures one necessarily comes
into the region where $\cd$ increases sharply. This has its origin in
coherent scattering from many impurities and ultimately gives rise to
localization.

One can finally notice the effects of different types of interactions on
localization. Roughly speaking, for forward scattering  repulsion ($g_2>0$)
enhances localization whereas attraction weakens it. In particular, strong
attraction leads to vanishing effective random potential. The delocalized
state then can be considered to be a true superconductor in the sense that
there is infinite conductivity even in an impure system. The effect of backward
scattering interactions is opposite to that of forward interactions.

\section{The Bethe ansatz: a pedestrian introduction}
The methods and results described in the preceding two sections were to
a large extent based on a continuum description of the physics of
one--dimensional electrons, and often relied on perturbation theory.
Complementary insight can be gained by considering models defined on a
lattice, and in particular some exactly solvable models. The physically
most important cases are the Heisenberg spin chain (the
subject of Bethe's original work) and its generalization for itinerant
fermions, the Hubbard model. In the following, I will discuss the exact
solution of the Heisenberg model in some detail, in the
hope to clarify the essential features. Subsequently, the necessary
generalizations for the Hubbard model will be indicated.

The exact solutions give
exact energies of the ground state and all the excited states in terms of
the solution of a system of coupled nonlinear equations. On the other hand,
the corresponding wavefunctions have a form so complicated that the explicit
calculation of matrix elements, correlation functions and other physical
quantities has remained impossible so far. In the subsequent sections
I shall describe
how the knowledge of the energy spectrum obtained from exact
solutions can be
combined with the results of the preceding two chapters to obtain a rather
detailed picture of the low--energy properties, in particular of correlation
functions.

It should be noticed that Bethe's method has been successfully applied
to a number of other interesting cases like the Kondo model
\cite{andrei_kondo_review,tsvelick_kondo_review} and 
also is by no means restricted to lattice
models.\cite{gaudin_fermions,yang_fermions,andrei_oNmodel,belavin_g1g2} For
more general and detailed discussions, the reader is referred to the
literature.\cite{bax_book,thacker_bethe_review,gaudin_book}

\subsection{The Heisenberg spin chain}
\subsubsection{Bethe's solution}
We will write the Hamiltonian of the Heisenberg spin chain of $L$
sites in the form
\begin{eqnarray}
\nonumber
H & = &  \sum_{i=1}^L (S ^x_i  S ^x_{i+1}
                 +  S ^y_i  S ^y_{i+1}
                 + \Delta S ^z_i  S ^z_{i+1}) \\
\label{eq:hh}
& = & \sum_{i=1}^L \left[\frac12(S ^+_i  S ^-_{i+1}
                 +  S ^-_i  S ^+_{i+1})
                 + \Delta S ^z_i  S ^z_{i+1} \right] \point
\end{eqnarray}
Here $\vec{S}_i=(S_i^x,S_i^y,S_i^z)$ is a spin--$1/2$ operator acting on
site
$i$,
$\delta$ is an anisotropy parameter that allows one to treat the
antiferromagnetic ($\Delta=1$), the ferromagnetic ($\Delta =
-1$), and general anisotropic cases\footnote{In the Hamiltonian,
eq.(\ref{eq:hh}), an overall sign opposite to that of some of the original
literature \cite{bethe_xxx,yang_xxz} is used}, and periodic boundary conditions
imply $\vec{S}_{L+1} = \vec{S}_1$. We notice that the Hamiltonian
conserves the $z$-- component of the total spin (for $|\Delta|=1$ total
spin is also conserved), and consequently one can write any
eigenfunction of a state with $N$ down and $L-N$ up spins as
\begin{equation}
| \psi \rangle = \sum_{x_1<x_2<\ldots<x_N} f(x_1,x_2,\ldots,x_N) |
x_1,x_2,\ldots,x_N \rangle \virg
\end{equation}
where the $x_i$ are the positions of the down spins. Applying the
Hamiltonian to this state, one obtains a ``Schr\"odinger equation'' for
the coefficients $f$. In particular, if none of the $x_i$ are adjacent,
one has
\begin{equation}
\label{eq:kin0}
\varepsilon f(\{x_i\}) = \frac12 \sum_{i=1}^N \delta_i f(\{x_i\}) \virg
\end{equation}
where $\varepsilon=E+\Delta(N-L/4)$, $E$ is the eigenenergy, and the
finite--difference operator $\delta_i$ is defined by
\begin{equation}\label{eq:diop}
\delta_i f(\ldots,x_i,\ldots) = f(\ldots,x_i-1,\ldots)+
 f(\ldots,x_i+1,\ldots) \point
\end{equation}
If two down spins are adjacent, e.g. $x_{k+1} = x_k+1$, displacement of
the spin is partly suppressed, and one finds instead of
eq.(\ref{eq:kin0})
\begin{eqnarray}
\nonumber
\varepsilon f(\ldots,x_k,x_k+1,\ldots) = \frac12 \sum_{i\neq
k,k+1}^N \delta_i f(\ldots,x_k,x_k+1,\ldots) \\
\nonumber
+\frac12 [f(\ldots,x_k-1,x_k+1,\ldots) +
f(\ldots,x_k,x_k+2,\ldots)] \\
\label{eq:kin1}
+\Delta f(\ldots,x_k,x_k+1,\ldots)
\end{eqnarray}
It should be quite clear how this generalizes to the case of more than
two $x_i$'s adjacent: whenever an down spin has an down neighbor, the term
representing
displacement to that site disappears, and one finds an ``interaction
term'' ($\Delta$) instead.

Bethe's solution starts by the observation that any linear combination
of plane waves of the form
\begin{equation}\label{eq:pw}
f(x_1,\ldots,x_N) = \sum_P A_P \exp \left[ i \sum_{j=1}^N k_{Pj}
x_j \right]
\end{equation}
is a solution of the ``free'' eq.(\ref{eq:kin0}) for arbitrary coefficients
$A_P$.  Here the summation is
over all permutations $P$ of the $N$ different wavenumbers $k_j$, i.e.
in
eq.(\ref{eq:pw}) the momenta are permuted amongst all the particles (down
spins).

The essential point in Bethe's solution is now to determine the coefficients
$A_P$ so that eqs.(\ref{eq:kin0}) and (\ref{eq:kin1}) become identical,
i.e. one requires
\begin{eqnarray} 
\nonumber
\lefteqn{2\Delta f(\ldots,x_k,x_k+1,\ldots) =} \\
\label{eq:cond}
&&  f(\ldots,x_k+1,x_k+1,\ldots) +
f(\ldots,x_k,x_k,\ldots) \point
\end{eqnarray}
Inserting this into eq.(\ref{eq:kin1}) one sees that one recovers
eq.(\ref{eq:kin0}) which is automatically solved by the plane wave form
(\ref{eq:pw}). One should notice in eq.(\ref{eq:cond}) that coefficients with two equal arguments
like $f(\ldots,x_k,x_k,\ldots)$ have no direct physical meaning but are of
course well--defined mathematically from eq.(\ref{eq:pw}).

To see how this works let us first consider the two--particle case. Then
there are only two permutations to sum over in eq.(\ref{eq:pw}), and
eq.(\ref{eq:cond}) becomes (after canceling common factors)
\begin{equation} \label{eq:p2}
2\Delta (A_0 \e^{i k_2} + A_1 \e^{i k_1}) =
(A_0+A_1) (1 + \e^{i (k_1+k_2)}) \point
\end{equation}
This then gives
\begin{equation} \label{eq:p3}
\frac{A_1}{A_0} = - \frac{1 + \e^{i(k_1+k_2)} - 2 \Delta \e^{i k_2}}{1 +%
\e^{i(k_1+k_2)} - 2 \Delta \e^{i k_1}}
= - \exp[-i\Theta(k_2,k_1)] \virg
\end{equation}
where the second equation defines the {\em phase shift} $\Theta$.
Note that this means that the plane--wave form of the solution,
eq.(\ref{eq:pw}), remains valid even when two down spins are at
nearest--neighbor sites where they interact. For longer range interactions
this would not in general be the case.

We now consider eq.(\ref{eq:cond}) for general $N$. It is then convenient to
associate with each permutation $P$ another permutation $P'$ which differs
from $P$ only by the exchange of two adjacent elements: $P'k=Pk+1, P'k+1=Pk$. One
then has
\begin{eqnarray}
\nonumber
\lefteqn{f(\ldots,x_k+m,x_k+n,\ldots)  =  {\sum_{P}}' \exp \left[i\sum_{j\neq
k,k+1}k_{Pj}x_j\right] \e^{i(k_{Pk}+k_{P'k})x_k}} \\
\label{eq:fxx2}
& & \times \left( A_P \e^{i( mk_{Pk}+nk_{P'k})} +
A_{P'} \e^{i( mk_{P'k}+nk_{Pk})} \right) \virg
\end{eqnarray}
where in eq.(\ref{eq:cond}) the cases $m,n=0,1$ are relevant, and the
summation is over half of the permutations (for example those with
$Pk<P'k$), the other permutations being
included explicitly by the term proportional to $A_{P'}$. Given that
eq.(\ref{eq:cond}) is supposed to be valid for any set $\{x_i\}$, the
coefficients of the plane wave factors in the left and right hand sides have
to be equal, e.g.
\begin{equation} \label{eq:p4}
2\Delta(A_P \e^{ik_{P'k}} + A_{P'} \e^{ik_{Pk}}) =
(A_P+A_{P'})(1+ \e^{i(k_{Pk}+k_{P'k})}) \point
\end{equation}
This is an obvious generalization of eq.(\ref{eq:p3}) and leads to
\begin{equation} \label{eq:p5}
\frac{A_{P}}{A_P'} = - \frac{1 + \e^{i(k_{Pk}+k_{P'k})} - 2 \Delta \e^{i k_{Pk}}}{1 +%
\e^{i(k_{Pk}+k_{P'k})} - 2 \Delta \e^{i k_{P'k}}}
= - \exp(-i \Theta(k_{Pk},k_{P'k})) \point
\end{equation}
This now fixes all the coefficients in eq.(\ref{eq:pw}) unambiguously, up to
an overall normalization, because every permutation can be build up from a
sequence of elementary permutation of two adjacent elements. To each such
 sequence
corresponds a product of factors like eq.(\ref{eq:p5}), one for each
elementary permutation, and moreover this product is independent of the
sequence of individual permutations, e.g.
\begin{eqnarray} 
\nonumber
&&(123) \rightarrow (213) \rightarrow (231) \rightarrow (321) \\
\label{eq:perm}
\mbox{and} &\;&
(123) \rightarrow (132) \rightarrow (312) \rightarrow (321)
\end{eqnarray}
lead to the same factor. One should also notice that eq.(\ref{eq:p5}) was
derived without any assumption on the other $x_i$'s, and consequently
situations with three or more adjacent down spins are equally covered and the
generalizations of eq.(\ref{eq:kin1}) to these cases are therefore also
fulfilled.

It remains to determine the allowed $k$--values. This follows from the periodic
boundary condition
\begin{equation} \label{eq:per}
f(1,x_2,\ldots,x_N) = f(x_2,\ldots,x_N,L+1) \point
\end{equation}
Inserting into eq.(\ref{eq:pw}) and noting that eq.(\ref{eq:per}) has to be
satisfied for all $(x_2,\ldots,x_N)$, one obtains the condition
\begin{equation} \label{eq:per2}
A_P/A_{P'} \e^{ik_{P'1}L} = 1 \virg
\end{equation}
which must be satisfied for every permutation $P$. Here $P'$ is a
permutation obtained from $P$ by a ``right shift'' of all elements:
\begin{equation} \label{eq:per3}
P' = (PN,P1,P2,\ldots,PN-1) \point
\end{equation}
The ratio of coefficients in eq.(\ref{eq:per2}) now can be calculated by
repeatedly permuting $PN$ in eq.(\ref{eq:per3}) to the right and using
eq.(\ref{eq:p5}). The resulting equation is
\begin{equation} \label{eq:per4}
(-1)^{N-1} \exp \left(i\sum_{j=1}^N \Theta(k_j,k_l) \right) \e^{ik_{l}L}=1
\virg
\end{equation}
with $\Theta$ defined in eq.(\ref{eq:p5}).

\subsubsection{The Heisenberg antiferromagnet}
We now turn to the particularly interesting and physically relevant case of
the Heisenberg antiferromagnet ($\Delta=1$). In this case, the fundamental
eq.(\ref{eq:per4}) can be transformed into a somewhat simpler form making
the transformation of variables
\begin{equation} \label{eq:tv}
\lambda = -\frac12 \cot (k/2) \Longleftrightarrow \e^{ik} =
\frac{2\lambda-i}{2\lambda+i} \point
\end{equation}
Equation (\ref{eq:per4}) then becomes
\begin{equation} \label{eq:lam1}
\left( \frac{2\lambda_l-i}{2\lambda_l+i} \right)^L = - \prod_{j=1}^N
\frac{\lambda_l-\lambda_j-i}{\lambda_l-\lambda_j+i} \virg
\end{equation}
and similarly one finds
the total momentum
\begin{equation} \label{eq:plam}
P = \sum_{i=j}^N k_j = \sum_{j=1}^N(2 \arctan(2\lambda_j)-\pi)
\end{equation}
and the energy
\begin{equation} \label{eq:elam}
E = L/4 + \sum_{j=1}^N(\cos k_j -1) = L/4- \sum_{j=1}^N
\frac{2}{1+4\lambda_j^2} \point
\end{equation}
The $z$--component of the total spin of the states obtained from
eq.(\ref{eq:lam1}) is obviously
$L/2-N$. In fact it can be shown that this is also the total spin of these
states, and that these states, together with those obtained by repeated
application of the lowering operator of the total spin, form the complete
set of eigenstates.

If all the $\lambda$'s are real, eq.(\ref{eq:lam1}) can be simplified by
taking the logarithm. One then obtains
\begin{equation} \label{eq:arc}
2L \arctan(2\lambda_l) = 2\pi J_l +2\sum_{j=1}^N
\arctan(\lambda_l-\lambda_j) \virg
\end{equation}
where the $J_l$ are integers if $L-N$ is odd and half--odd integers (of the
form 1/2, 3/2, ...) is $L-N$ is even and are restricted to
\begin{equation} \label{eq:res}
|J_l| < (L-N+1)/2 \point
\end{equation}
In the ground state one has $N=L/2$, and then all the $J_l$ are uniquely
determined. In the thermodynamic limit, the allowed $\lambda$'s are very
close to each other, and their distribution then can be determined by a
linear integral equation. The ground state energy then is $E_0 = L(1/4 - \ln
2)$.

\begin{figure}[htb]
\centerline{\epsfxsize 6cm
{\epsffile{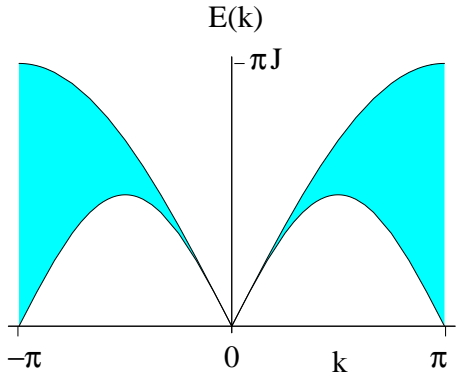}}
}
\caption{The continuum of low--lying triplet and singlet states. The lower
and upper limit of the continuum are $(\pi/2) |\sin k|$ and $\pi
|\sin(k/2)|$,
respectively.}
\label{f6:1}
\end{figure}

\begin{figure}[htb]
\centerline{\epsfxsize 7cm
\epsffile{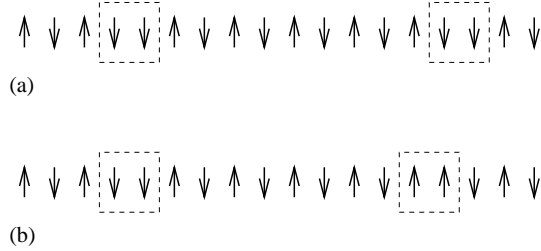}
}
\caption{Two--spinon configurations in an antiferromagnetic chain with
$m_z=0$ (a) and $m_z=1$ (b).}
\label{f6:2}
\end{figure}

Excited states can be obtained by taking out one of the down spins. One then
has a state of total spin $1$. One now has to choose $N=L/2-1$, subject to
the constraint (\ref{eq:res}), i.e. one has to take two (half--)integers
out of the sequence $-L/4...L/4$. This is thus a two--parameter family of
states, which, in the thermodynamic limit has energy
\begin{equation} \label{eq:}
E(p_1,p_2) = \varepsilon(p_1) + \varepsilon(p_2) \virg \quad \mbox{with}
\quad \varepsilon(p) = \frac{\pi}{2} \sin p
\end{equation}
and momentum $P=p_1+p_2$. At fixed total momentum $P$ there is thus {\em a
continuum} of allowed states, as shown in fig.\ref{f6:1}. This energy--momentum relation suggests that the
excitation is actually the combination of {\em two} elementary excitations
which don't interact, so that there energies and momenta just add up.  This
interpretation finds further support in the fact that excitations with total
spin 0 exist that have the same energetics as the spin 1 excitations (these
correspond to the appearance of one pair of complex conjugate $\lambda$'s in
eq.(\ref{eq:lam1})). These singlet and triplet excitations thus can in fact
be seen as excitations of pairs of spin $1/2$ objects, called ``spinons'',
which do not interact but form states of total spin 0 or 1. Spinons can be
visualized as kinks in an antiferromagnetic order parameter, as shown in
fig.\ref{f6:2}. In an
isotropic Heisenberg chain, these individual spinons get delocalized into
plane wave states due to the exchange interaction.  For an even number of
sites the total spin is always integer, so that the spinons always have to
be excited pairwise. However, once created, they behave essentially as
noninteracting objects.

In the Heisenberg model, the spinons are only noninteracting in the
thermodynamic limit $L\rightarrow \infty$. However, in a modified model with
$1/r^2$ interaction, this separation is true for all states even in a finite
system \cite{haldane_inv_square,shastry_inv_square}. One can further show
that individual spinons behave as semions, i.e. their statistical properties
are in many respects intermediate between fermions and
bosons\cite{haldane_stat}.

\subsubsection{The Jordan--Wigner transformation and spinless fermions}
The spin model discussed above can be transformed into a model of spinless
fermions, noting that $S_i^+$ and $S_i^-$ anticommute. The {\em
Jordan--Wigner transformation} \cite{jor_tran} then relates spin to fermion
operators ($a^{\phantom{\dagger}}_i, a^\dagger_i$) via
\begin{equation} \label{eq:jw}
S_i^+ = a_i^\dagger \exp \left[i\pi\sum_{j=1}^{i-1} a^\dagger_j a^{\phantom{\dagger}}_j \right] \virg
S_i^z = a^\dagger_i a^{\phantom{\dagger}}_i - \frac12 \point
\end{equation}
Presence or absence of a fermion now represent an up or down spin, and the
exponential factor insures that spin operators on different sites commute,
whereas fermionic operators of course anticommute. The transformation can
now be used to rewrite the spin Hamiltonian (\ref{eq:hh}) in terms of
fermions as
\begin{equation} \label{eq:hhfer}
H = \sum_{i=1}^L \left[ \frac12(a^\dagger_i a^{\phantom{\dagger}}_{i+1} + a^\dagger_{i+1} a^{\phantom{\dagger}}_i) + \Delta
(a^\dagger_ia^{\phantom{\dagger}}_i -\frac12)(a^\dagger_{i+1} a^{\phantom{\dagger}}_{i+1} -\frac12) \right] \point
\end{equation}
The spin model thus is transformed into a fermionic model, with the
``spin--flip'' terms giving rise to motion of the fermions, whereas the
$S^z$--$S^z$ interaction gives rise to a fermion--fermion interaction
between adjacent sites.

The fermionic analogy now allows one to study correlation functions of the
spin chain using the bosonization method developed in section
\ref{sec:bosonspl}. In particular, the spin--spin correlations for the
isotropic antiferromagnet decays as \cite{luther_chaine_xxz}
\begin{equation} \label{eq:spincorr}
\langle \vec{S}_i \cdot \vec{S}_j \rangle \approx (-1)^{|i-j|}\frac{\ln^{1/2}
|i-j|}{|i-j|} \virg
\end{equation}
where the logarithmic corrections term comes from contributions of a
marginally irrelevant umklapp interaction term
\cite{affleck_log_corr,giamarchi_logs,singh_logs}. The alternation in
eq.(\ref{eq:spincorr}) indicates the expected tendency towards
antiferromagnetic order, but correlations do decay (with a rather slow power
law), and there thus is no long--range order, as to be expected in one
dimension. The analysis can be extended to give the full dynamic structure
factor \cite{schu_spins}, and recent experiments on $\rm KCuF_3$ confirm
these results.\cite{tennant_kcuf}

For $\Delta < 1$, a similar law holds, but with a $\Delta$--dependent
exponent and no logarithmic term. On the other hand, for $\Delta >1$ the
spins are preferentially aligned along the $z$--direction, and one then has
a long--range ordered ground state of the Ising type. There thus is a phase
transition exactly at the isotropic point $\Delta=1$. In the fermionic
language, this corresponds to a metal--insulator transition
\cite{shankar_spinless}.

\subsection{The Hubbard model}
\label{hubsec}
The Hubbard model is the prototypical model used for the description of
correlated fermions in a large variety of circumstances, ranging from
high--$\rm T_c$ superconductors to heavy fermion compounds and organic
conductors. In spite of its apparent simplicity, there is still no general
solution, or even a consensus on its fundamental properties. Notable
exceptions are the cases of one and infinite dimensions
\cite{vollhardt_jeru,georges_infdim}. In particular,
in one dimension an exact solution is
available.\cite{lieb_hubbard_exact}

The Hamiltonian in one dimension has the well--known form
\begin{equation}
H = -t \sum_{i,s} (a_{i,s}^\dagger a^{\phantom{\dagger}}_{i+1,s} + a_{i+1,s}^\dagger a^{\phantom{\dagger}}_{i,s})
+ U \sum_i n_{i,\uparrow} n_{i,\downarrow} \virg
\end{equation}
where  $n_{i,s}$ is the  number operator for fermions of spin $s$, and
the sum is over the
$L$ sites of a one--dimensional chain with periodic boundary conditions.

The model has {\em two} global SU(2) symmetries
\cite{pernici_su2,zhang_su2,schulz_su2}:
the first is the well--known spin rotation invariance and the second is
particular to the Hubbard model and relates sectors of different particle
numbers.  The total symmetry thus is $SU(2) \times SU(2) \simeq SO(4)$. One
should notice that more complicated interactions, e.g. involving further
neighbors, will conserve the spin rotation invariance but in general not the
``charge'' SU(2) invariance. Rather, this second symmetry will become the
standard global $U(1)$ invariance associated with particle number
conservation.

\subsubsection{The exact solution}
\label{exsec}
The exact solution of the one--dimensional Hubbard model has been found by
Lieb and Wu \cite{lieb_hubbard_exact}.
To obtain the wavefunctions of the Hubbard model, one writes a
general $N$--particle state as
\begin{equation}
| F \rangle = \sum_{x_1} \ldots \sum_{x_N} \sum_{s_1} \ldots \sum_{s_N}
F_{s_1,\ldots,s_N}(x_1,\ldots,x_N) \prod_{i=1}^N a_{x_i,s_i}^\dagger |vac\rangle
\point
\end{equation}
The simplest case is the two--particle problem. Then in the two parts of
configuration space $x_2 > x_1$ (region I) and $x_1 > x_2$ (region II) the
wavefunction
is just a product of two plane waves, and the only nontrivial effects occur
for $x_1= x_2$.
The full wavefunction $F$ then is
\begin{equation}
F_{s_1,s_2} = e^{i(k_1x_1 + k_2x_2)} [ \theta(x_2- x_1) \xi^I_{s_1,s_2} +
\theta(x_1- x_2) \xi^{II}_{s_1,s_2} ] \virg
\end{equation}
where the factors $\xi$ depend only on the spin quantum numbers of the two
particles, and $\theta$ is the usual step function, with $\theta(0)=1/2$.
In order to satisfy the Schr\"odinger equation, these
coefficients have to obey the equation
\begin{equation}
\xi^{II}_{s_1,s_2} = \sum_{t_1,t_2} S_{s_1s_2,t_tt_2} \xi^I_{t_1t_2}
\point
\end{equation}
The {\em S--matrix} has the form
\begin{equation}  \label{smat}
S_{\alpha\beta,\gamma\delta} =
\frac{t(\sin k_1 - \sin k_2) + iUP_{\alpha\beta,\gamma\delta}}%
{t(\sin k_1 - \sin k_2 ) + iU}
\virg
\end{equation}
where $P_{\alpha\beta,\gamma\delta}$ is the operator permuting two spins.
This operator acts on the spin part of the two--particle Hilbert space.

For $N$ particles, similar to the Heisenberg case, one has matching
conditions whenever two particles cross. The major complication in the
Hubbard model comes from the fact that these matching conditions obviously
involve the spin of the two particles and therefore involve the S--matrix,
eq.(\ref{smat}), rather than just complex phase factors (compare with
eq.(\ref{eq:p3})). In order to fulfill compatibility conditions for
different paths in configuration space like those of eq.(\ref{eq:perm}) one
has has to have
\begin{equation}
S^{23}S^{13}S^{12} = S^{12}S^{13}S^{23} \point
\end{equation}
These are the famous {\em Yang--Baxter equations} \cite{yang_fermions} which
have to be satisfied
for a system to be solvable by Bethe Ansatz. One can verify that these
equations are indeed satisfied by the S--matrix of the Hubbard model, eq.
(\ref{smat}).
The subsequent analysis is based on a ``generalized
Bethe ansatz''
\cite{gaudin_fermions,yang_fermions}. A
detailed description of the methods used can be found in the specialized
literature \cite{bax_book,thacker_bethe_review}, and a very detailed
derivation is given in  Gaudin's book \cite{gaudin_book}.

Imposing periodic boundary conditions, the allowed values of $k_j$ are
obtained from the solution of the coupled set of nonlinear equations
\begin{eqnarray}                      \label{be1}
e^{ik_jL} & = & \prod_{\alpha = 1}^M e\left(\frac{4(\sin k_j -
\lambda_\alpha)}{U} \right) \\
\label{be2}
\prod_{j=1}^N e\left( \frac{4(\lambda_\alpha - \sin k_j)}{U} \right) & = &
- \prod_{\beta=1}^M e \left( \frac{2(\lambda_\alpha - \lambda_\beta)}{U}
\right) \virg
\end{eqnarray}
Here $N$ is the total number of electrons, $M$ is the number of down--spin
electrons ($M \le N/2$), and $e(x) = (x+i)/(x-i)$.  The $\lambda_\alpha$ are
parameters characterizing the spin dynamics. We note that in general, both
the $k_j$'s and the $\lambda$'s are allowed to be complex. The energy and
momentum of a state are
\begin{equation}
E = -2t \sum_{j=1}^N \cos k_j \quad \virg \quad \quad P = \sum_{j=1}^N k_j
\point
\end{equation}

The determination of all the solutions of eqs.(\ref{be1}, \ref{be2}) is not
easy. It has recently been shown that under certain assumptions these
equations do indeed give all the ``lowest weight'' (with respect to $SU(2)
\times SU(2)$) eigenstates of the Hubbard model.  The complete set of
eigenstates then is obtained acting repeatedly with raising operators on the
Bethe ansatz states.\cite{essler_hubbard_complete}

In the ground state all the $k$'s and $\lambda$'s are real.
Numerical results for the ground state energy as a function of particle
density and $U$ have been given by Shiba \cite{shiba_hubbard_exact}.
The excited states are discussed in some detail in
the literature \cite{shiba_hubbard_exact,coll_excitations,woynarovich_spin,%
woynarovich_complexk,schulz_jeru}. I here just briefly summarize the
different types:
\begin{enumerate}
\item ``$4\kf$'' singlet states.
The excitation energy goes to zero
at $q=0$ and $q=4\kf$.
These excitations form a two--parameter continuum and only exist away
from half--filling.
\item ``$2k_F$'' singlet and triplet states.
These are degenerate two--parameter families of
solutions, similar to the Heisenberg chain. The energy goes to zero for $q=0$
and $q=2\kf$.
\item There are also states with complex $k$. The energy of these states is
proportional to $U$ for large $U:t$, i.e. some sites are doubly occupied
(``upper Hubbard band'').
\end{enumerate}
In addition, there are of course states with added particles or holes.

For $U \rightarrow \infty$ the ratio $\sin k_j/U$ in eqs.(\ref{be1},\ref{be2})
clearly vanishes, however there is no restriction on the $\lambda$'s.
Introducing the scaled variables $\Lambda_\alpha = 2 \lambda_\alpha /U$, the
``spin equation'' (\ref{be2}) becomes
\begin{equation}
\left(\frac{2 \Lambda_\alpha + i}{2 \Lambda_\alpha - i} \right)^N
=
- \prod_{\beta=1}^M
\frac{\Lambda_\alpha - \Lambda_\beta +i}{\Lambda_\alpha - \Lambda_\beta -i}
\virg
\end{equation}
which are the well--known Bethe ansatz equations for the spin--1/2
Heisenberg chain, eq.(\ref{eq:lam1}), i.e. the spin wavefunction
of
$N$ particles is just that of an $N$--site (not $L$--site!) Heisenberg
chain, even when
there is less than one particle per site.

\subsubsection{Luttinger liquid parameters}
\label{lutt}
The above results provide a complete picture of the energy spectrum of
the one--dimensional Hubbard model (and also of its thermodynamics
\cite{hubbard_thermo}). On the other hand, the wavefunctions
obtained are prohibitively complicated, and no practical scheme for the
evaluation of expectation values or correlation functions has been
found. Of course,
in a weakly interacting system the coefficients
$K_\rho$ and $u_\nu $ can be determined
perturbatively. For example, for the Hubbard model
one finds
\begin{equation}  \label{pert1}
K_\rho = 1 -  U/(\pi v_F) + ... \virg
\end{equation}
where $ v_F = 2t \sin(\pi n/2)$ is the Fermi velocity for $n$ particles per
site.  For larger $U$ higher operators appear in the continuum Hamiltonian
(\ref{hbos}), e.g. higher derivatives of the fields or cosines of multiples
of $\sqrt8 \phi_\sigma$.  These operators are irrelevant, i.e. they
renormalize to zero and do not qualitatively change the long-distance
properties, but they do lead to nontrivial corrections to the coefficients
$u_\nu,K_\rho$.  In principle these corrections can be treated order by
order in perturbation theory. However, this approach is obviously
unpractical for large $U$, and moreover it is quite possible that
perturbation theory is not convergent.  To obtain the physical properties
for arbitrary $U$ a different approach \cite{schulz_hubbard_exact} can be
used which will now be explained.

I note two points: (i) in the small-$U$ perturbative regime, interactions
renormalize to the weak-coupling fixed point $g_1^*=0, K_\sigma^* =1$; (ii)
the exact solution \cite{lieb_hubbard_exact} does not show any singular
behavior at nonzero $U$, i.e. large $U$ and small $U$ are the same phase of
the model, so that the long-range behavior even of the large $U$ case is
determined by the fixed point $g_1^* = 0$. Thus, the low energy properties
of the model are still determined by the three parameters $u_{\rho,\sigma}$
and $K_\rho$.
\begin{figure}[htbp]
\centerline{{\epsfxsize=7cm  \epsffile{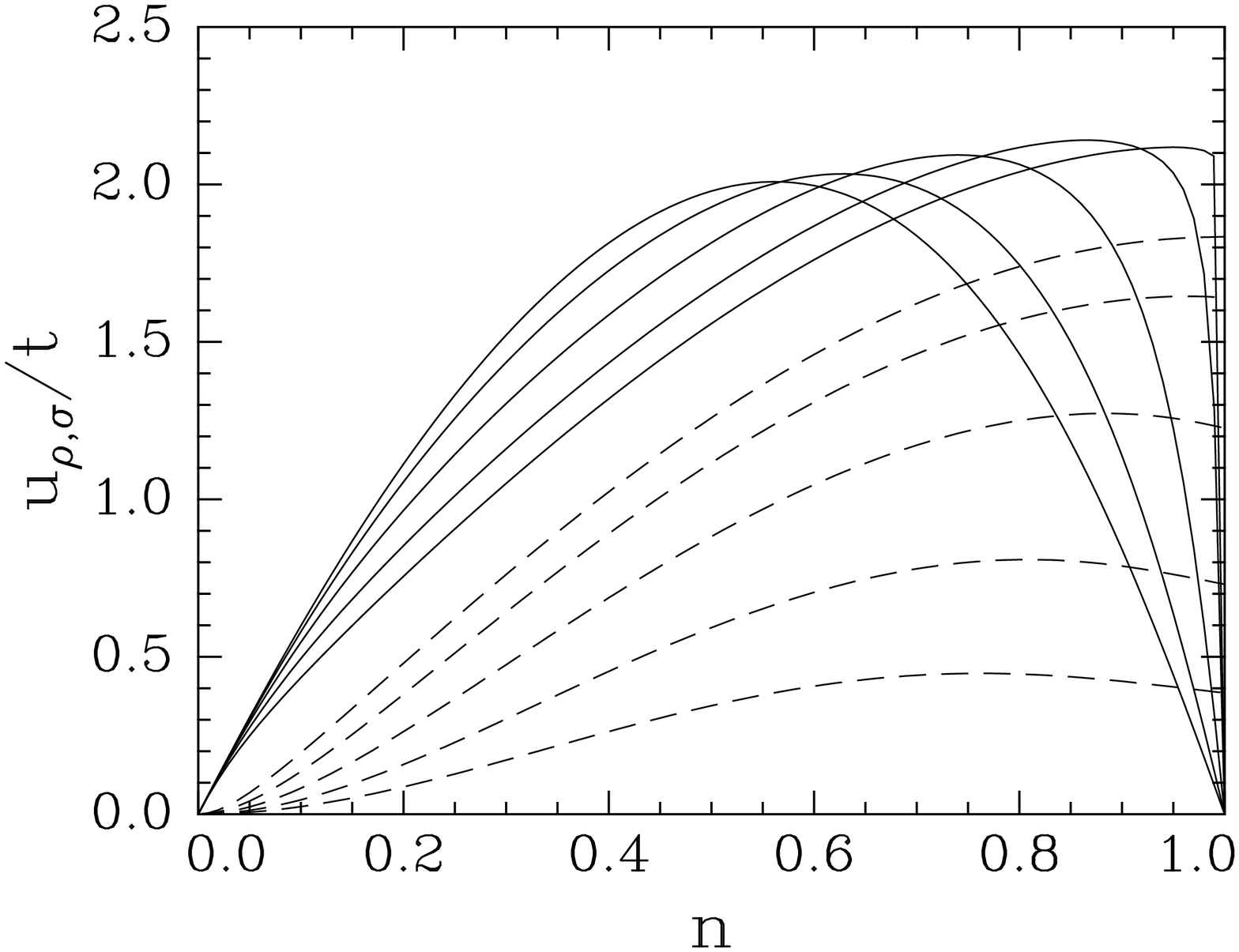}}}
\vspace*{-0.5cm}
\caption[dummy]{\em The charge and spin velocities $u_\rho$ (full
line) and $u_\sigma$ (dashed line)
for the Hubbard model, as a function of the band filling for different
values of $U/t$: for $u_\sigma$ $U/t=1,2,4,8,16$ from top to bottom,
 for $u_\rho$ $U/t=16,8,4,2,1$ from top to bottom in the left part of
the figure. \label{ur}
}
\end{figure}

The velocities $u_{\rho,\sigma}$ can be obtained from the long wavelength
limit of the excitations discussed above. In particular, the $4k_F$
component of the charge correlation function only involves the $\phi_\rho$
field, and it is therefore natural to obtain $u_\rho$ from the $q\rightarrow
0$ limit of the ``$4k_F$'' singlets. Similarly, $u_\sigma$ is obtained from
the ``$2k_F$'' excitations.  In the thermodynamic limit the corresponding
excitation energies are easily found from the numerical solution of a linear
integral equation.\cite{coll_excitations} Results are shown in \fref{ur} for
various values of $U/t$. Note that for $U=0$ one has $u_\rho = u_\sigma = 2t
\sin (\pi n/2)$, whereas for $U \rightarrow \infty$ $u_\rho = 2t \sin (\pi
n)$, $u_\sigma = (2 \pi t^2 /U) (1 -\sin (2\pi n)/(2\pi n))$. In the
noninteracting case $u_\sigma \propto n$ for small $n$, but for {\em any}
positive $U$ $u_\sigma \propto n^2$.

\begin{figure}[htbp]
\centerline{\epsfxsize=7cm  \epsffile{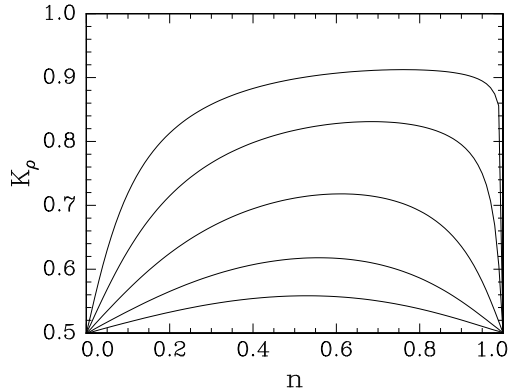}}
\caption[dummy]{\em
The correlation exponent $K_\rho$
as a function of the bandfilling $n$ for different
values of $U$ ($U/t = 1,2,4,8,16$ for the top to bottom curves).
Note the rapid variation near $n=1$ for small $U$.}
\label{krho}
\end{figure}

To obtain the parameter $K_\rho$ from the exact solution note that the
gradient of the phase field $\phi_\rho$ is proportional to the particle
density, and in particular a constant slope of $\phi_\rho$ represents a
change of total particle number. Consequently, the coefficient $u_\rho /
K_\rho $ in eq. (\ref{hnu}) is proportional to the variation of the ground
state energy $E_0$ with particle number:
\begin{equation}    \label{kr}
\frac{1}{L} \frac{\partial^2 E_0(n)}{\partial n^2} = \frac{\pi}{2}
\frac{u_\rho}{K_\rho} =  \frac{1}{n^2\kappa} \point
\end{equation}
Equation (\ref{kr}) now allows the direct determination of $K_\rho$:
$E_0(n)$ can be obtained solving (numerically) Lieb and Wu's
\cite{lieb_hubbard_exact} integral equation, and $u_\rho$ is already
known. The results for $K_\rho$ as a function of particle density are shown
in
\fref{krho}{ } for different values of $U/t$.
For small $U$ one finds in all cases agreement with the perturbative
expression, eq. (\ref{pert1}), whereas for large $U$ $K_\rho \rightarrow
1/2$.  The limiting behavior for large $U$ can be understood noting that for
$U=\infty $ the charge dynamics of the system can be described by
noninteracting {\em spinless} fermions (the hard-core constraint then is
satisfied by the Pauli principle) with $k_F$ replaced by
$2k_F$. Consequently one finds a contribution proportional to $ \cos (4k_F
x) x^{-2}$ in the density-density correlation function, which from
eq. (\ref{nn}) implies $K_\rho = 1/2$. One then finds an asymptotic decay
like $\cos(2k_Fx) x^{-3/2} \ln^{1/2}(x)$ for the spin-spin correlations,
eq.(\ref{ss}), and an exponent $\alpha=1/8$ in the momentum distribution
function. The result $\alpha = 1/8$ has also been found by Anderson and
Ren,\cite{anderson_ren_losal} and by Parola and Sorella
\cite{parola_inf}. Ogata and Shiba's
numerical results \cite{ogata_inf} are quite close to these exact values.

We note that in the whole parameter region, as long as the interaction is
repulsive one always has $K_\rho < 1$, which means that magnetic
fluctuations are enhanced over the noninteracting case. On the other hand,
superconducting pairing is always suppressed.

It should be emphasized here that the results of \fref{krho} are valid
for $n \rightarrow 1$, but not for $n=1$. In fact, in that latter case,
there is a gap in the charge excitation spectrum, as expected from the
umklapp term discussed below, and the correlations of $\phi_\rho$ become long
ranged. Close to half--filling, the asymptotic behavior of the charge part
of correlation functions like (\ref{nn}) is essentially determined by the
motion of the holes. Writing the density of holes as $\rho = 1/(1-n)$ one
then expects a crossover of the form \cite{schulz_cic2d}
\begin{equation}      \label{cross}
\langle n(x) n(0) \rangle \approx \cos (2k_Fx) [1 + (\rho x)^2
]^{K_\rho/2} x^{-1} \ln^{-3/2}(x)
\end{equation}
for the $2k_F$ part of the density correlation function, and similarly for
other correlation functions. Clearly, only for $x \gg 1/\rho$ are the
asymptotic power laws valid, whereas at intermediate distances $ 1 \ll x \ll
1/\rho$ one has effectively $K_\rho = 0$. Clearly, the form (\ref{cross})
provides a smooth crossover as $n \rightarrow 1$.

Results equivalent to the present ones can be obtained using the conformal
invariance of the Hubbard model \cite{frahm_confinv,kawakami_hubbard}.
These results have subsequently be generalized to the case with an applied
magnetic field \cite{frahm_confinv_field}. The method for calculating the
exponent $K_\rho$ is quite general and has been used for example for
generalized Hubbard models \cite{schulz_hubbard_exact,mila_zotos} and the
t--J model.\cite{kawakami_tj,ogata_tj}

\subsubsection{Transport properties and the metal--insulator transition}
\begin{figure}[htbp]
\centerline{{\epsfxsize=7cm  \epsffile{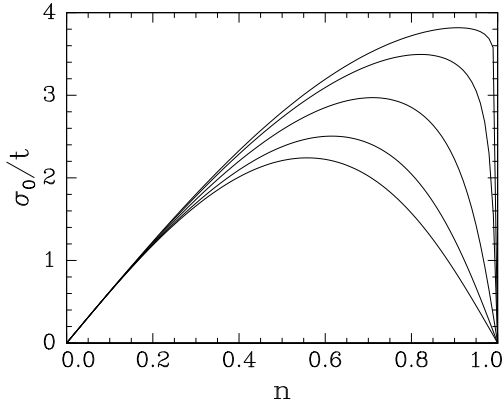}}}
\caption[dummy]{\em
The weight of the dc peak in $\sigma (\omega )$ as a
function of bandfilling for different values of $U/t$
($U/t = 1,2,4,8,16$ for the top to bottom curves).
}
\label{sig}
\end{figure}

The exact solution of Lieb and Wu can also be combined with the
long--wavelength effective Hamiltonian (\ref{hbos}) to obtain some
information on the frequency--dependent conductivity $\sigma (\omega)$.  In
particular, from eq. (\ref{sig0}) there is a delta function peak at zero
frequency of weight $\sigma_0=2 K_\rho u_\rho$, with the results plotted in
\fref{sig}. As expected, far from half--filling, the dc weight is nearly
independent of interaction strength.  peak. On the other hand, for
$n\rightarrow 1$ $\sigma_0$ vanishes linearly, implying a linear variation
of the ratio $n / m^*$ with ``doping'', and for exactly half--filling the dc
conductivity vanishes. There thus is a {\em metal--insulator transition} at
$n=1$, with the insulating state existing only at $n=1$.
One can also obtain the total frequency--integrated
conductivity.\cite{schulz_hubbard_exact} One then sees that as
$n\rightarrow 1$ more and more weight is transferred from the
zero--frequency peak to higher frequencies, essentially but not
exclusively to frequencies above a ``Mott--Hubbard gap''.

It is interesting to discuss the metal--insulator transition in more
detail. In the bosonization description, the insulating state at half
filling is due to an extra term
\begin{equation}\label{eq:hu}
H_u = \frac{2g_3}{(2\pi\alpha)^2} \int dx \cos(\sqrt8 \phi_\rho)
\end{equation}
which is due to umklapp scattering.
The charge part of the Hamiltonian can then be transformed
into a model of massive fermions, with energy--momentum relation
$\varepsilon_k = \pm (v^2 k^2 + \Delta_c^2)^{1/2}$,%
\cite{emery_etal_umklapp} where $\Delta$ is the charge excitation gap
created by umklapp scattering.  At half--filling, this term leads to a gap
$\Delta_c$ in the charge excitation spectrum and therefore to insulating
behavior.  At half-filling all negative energy states are filled, all
positive energy states are empty, and one has an insulator.  Doping with a
concentration $n^*$ of holes ($n<1$), some of the negative energy states
become empty and only states with $|k| > k_F^* \propto n^*$ are
filled. Because of the vanishing interaction, a standard formula can be used
\cite{chaikin_thermopower} and gives a positive thermopower, i.e. {\em
approaching the metal--insulator transition from $n < 1$, the thermopower is
hole--like},\cite{schulz_trieste,stafford_thermop} whereas obviously far
from the transition ($n \ll 1$) it is electron--like. The exactly opposite
behavior can be found for $n > 1$.

The fact
that $u_\rho$ and $\sigma_0$ vanish linearly as $n \rightarrow 1$ seems
to be consistent with a divergent effective mass at constant carrier
density because $u_\rho \approx 1/m^*, \sigma_0 \approx n/m^*$. A
constant carrier density is also consistent with the fact that $k_F =
\pi n /2$ is independent of $U$. It is {\em not consistent} with the
hole--like sign of the thermopower as $n \rightarrow 1$ from below, nor
with the electron--like sign as $n \rightarrow 1$ from above: if the
carriers are holes, the carrier density is the density of holes: $n^* =
1-n$. Treating the holes as spinless fermions, as already mentioned
before, one expects $\sigma_0 \rightarrow 0$ because $n^*
\rightarrow 0$, and $\gamma \rightarrow \infty$ because the density of
states of a one--dimensional band diverges at the band edges.
This agrees with what was found explicitly in section \ref{lutt}.
What is not so easily understood in this picture is the fact that $k_F$
(i.e. the location of the singularity of $n_k$) is given by its
free--electron value $\pi n /2$, rather then being proportional to
$n^*$. One should however notice that $n_k$ is given by the
single--particle Green's function, which contains both charge and spin
degrees of freedom. The location of $k_F$ then may possibly be explained
by phase
shifts due to holon--spinon interaction. This is in fact suggested by
the structure of the wavefunction of the exact solution
\cite{ogata_inf}.

The magnetic properties do not agree with what one expects from an effective
mass diverging as $n \rightarrow 1$: $u_\sigma$ and therefor $\chi$ remain
finite. Moreover, the NMR relaxation rate would have the behavior $1/T_1 =
\alpha T + \beta \sqrt{T}$, where the first (Korringa) term comes from
fluctuations with $q \approx 0$, whereas the second term comes from
antiferromagnetic fluctuations with $q \approx 2k_F$. None of these
properties is strongly influenced by the diverging effective mass observed
e.g. in the specific heat. This fact is of course a manifestation of the
separation between spin and charge degrees of freedom.

The discussion can be generalized to metal--insulator transitions near
other rational bandfillings, i.e. one--third or one--quarter filled
bands.\cite{schulz_la} In the Hubbard model, however, an insulating
phase occurs only at half--filling.

It is worthwhile to point out here that the behavior of the one--dimensional
system is quite different from that of scenarios for the metal--insulator
transition in Fermi liquids. For example, in the ``nearly--localized''
description,\cite{vollhardt_helium_revue} effective mass effects dominate,
giving rise to a simultaneous divergence of specific heat and spin
susceptibility. On the other hand, in infinite dimension
\cite{vollhardt_jeru,georges_infdim} a behavior similar
to the one--dimensional case is observed: approaching the insulating state by
varying the bandfilling the spin susceptibility remains finite but the
specific heat coefficient diverges.\cite{georges_private}

\subsubsection{Spin--charge separation}
The Hubbard model also provides a rather straightforward interpretation
of the spin--charge separation discussed above. Consider a piece of a
Hubbard chain with a half--filled band. Then for strong $U$ there will
be no doubly--occupied sites, and because of the strong short--range
antiferromagnetic order the typical local configuration will be
$$
\begin{array}{c}
\cdots
\uparrow \downarrow \uparrow \downarrow \uparrow \downarrow
\uparrow \downarrow \uparrow \downarrow \uparrow \downarrow
\cdots
\end{array}
$$
Introducing a hole will lead to
$$
\begin{array}{c}
\cdots
\uparrow \downarrow \uparrow \downarrow \uparrow O
\uparrow \downarrow \uparrow \downarrow \uparrow \downarrow
\cdots
\end{array}
$$
and after moving the hole one has (note that the kinetic term in the
Hamiltonian does not flip spins)
$$
\begin{array}{c}
\cdots
\uparrow \downarrow O \uparrow \downarrow \uparrow
\uparrow \downarrow \uparrow \downarrow \uparrow \downarrow
\cdots
\end{array}
$$
Now the hole is surrounded by one up and one down spin,
whereas somewhere else there are two adjacent up spins. Finally, a few
exchange spin processes lead to
$$
\begin{array}{c}
\cdots
\uparrow \downarrow O \uparrow \downarrow \uparrow
\downarrow \uparrow \downarrow \uparrow \uparrow \downarrow
\cdots
\end{array}
$$
Note that the original configuration, a hole surrounded by {\em two} up
spins has split into a hole surrounded by antiferromagnetically
aligned spins (``holon'') and a domain--wall like configuration, two
adjacent up spins, which
contain an excess spin 1/2 with respect to the initial antiferromagnet
(``spinon'').
The exact solution by Lieb and Wu contains two types of
quantum numbers which
can be associated with the dynamics of the spinons and holons,
respectively.
We thus notice that spinons and holons
\cite{kivelson_holon,zou_holon} have a well-defined meaning in the
present one--dimensional case.

The above pictures suggest that, as far as charge motion is concerned,
the Hubbard model away from half--filling can be considered as a
one--dimensional harmonic solid, the motion of the holes providing for an
effective elastic coupling between adjacent electrons. This picture has been
shown to lead to the correct long--distance correlation functions for
spinless fermions
\cite{haldane_bosons,emery_jeru}. For the case with spin, this suggests
that one can consider the system as a harmonic solid with a spin at each
site of the elastic lattice (lattice site = electron in this picture), and
this gives indeed the correct spin correlations \cite{schulz_jeru}.

\section{Conclusion and outlook}
In these notes, the different behavior of interacting many--fermion systems
in one and higher dimension have been contrasted. In dimension bigger than
one, Landau's Fermi liquid theory is based on the postulated existence of 
quasiparticles which have properties closely related to those of a
noninteracting system. On the other hand in one dimension, essentially
because of the geometric constraints on the particle dynamics no well--defined
quasiparticle excitations exist. Instead, the low--lying excitations are
collective spin or charge density oscillations, giving rise to the
interesting phenomenon of spin--charge separation, as well as to
non--universal power law behavior of correlation functions. This {\em
Luttinger liquid} physics is most appropriately described by the
bosonization formalism developed in sec.\ref{wcsec}. We may note here that
bosonization can equally be applied in higher
dimension,\cite{houghton_bos_3d,fradkin_bos_3d} however at least for
``normal'' interactions one finds Fermi liquid properties.

It is clearly of interest to investigate the possibility of a crossover
between Luttinger and Fermi liquid. A most natural candidate for such
behavior are systems of parallel chains coupled by a small interchain
hooping integral $t_\perp$. Such a model is appropriate for the discussion or
real {\em quasi--}one-dimensional conductors. This type of models has been
studied for some time but there is still a number of open
questions. Nevertheless, most work points to a crossover between Luttinger
(at high temperature) and Fermi liquid (at low temperature) behavior at a
temperature determined by $t_\perp$.
\cite{prigodin_firsov,brazov_transhop,bourbon_couplage,boies_couplage,%
schulz_2chain}
At lower temperatures three--dimensionally ordered phases can appear. A
different scenario has been advocated by Anderson and collaborators
\cite{anderson_q1d,clarke_couplage} who argue that even well below $T\approx
t_\perp$ particle movement in the direction transverse to the chains
remains incoherent, i.e. there would still be no three--dimensional Fermi
liquid. Another approach is to consider the spatial dimension as a
continuously varying parameter.\cite{castellani_continuousD} One then finds
that the Luttinger liquid disappear as soon as $d>1$.  

It is worthwhile to mention here that the question of Fermi liquid or
non--Fermi liquid is of interest in a number of other contexts. For example,
the low--temperature properties of a ``normal'' Kondo impurity can be
described as a local Fermi liquid.\cite{nozieres_kondo} On the other hand, the
so--called ``overscreened'' Kondo effect \cite{nozieres_blandin} is
characterized by non--Fermi liquid behavior, in particular a divergent local
spin susceptibility.\cite{affleck_kondo_over} A review of this subject has been given by
Emery and Kivelson.\cite{emery_kivelson_kondo_review}

The other major area of current interest in possible non--Fermi liquid
behavior is the physics of strongly correlated systems in two dimensions,
and in particular of the high--T$_c$ superconducting copper oxides. Here a
number of experimental results in the normal conducting state suggest
non--Fermi liquid
behavior.\cite{anderson_hgtc_hubbard,anderson_luttinger,varma_marginal} 
The theoretical situation is much less clear. On the one hand, a number of
radically non--Fermi liquid pictures have been proposed, e.g. a
two--dimensional Luttinger liquid \cite{anderson_luttinger} or the
so--called gauge theory approach.\cite{nagaosa_lee} On the other hand,
there are also models with only relatively minor modifications of the Fermi
liquid picture. A rather complete overview of the rather unclear current
situation (1994) can be found in recent conference proceedings.\cite{hgtc_94}  
Finally, an interesting non--Fermi liquid state has been proposed for the
half--filled Landau level in two dimensions.\cite{halperin_lee_read}

\section*{Acknowledgments}
My understanding of the subjects exposed here has greatly benefitted from
numerous discussions with J. Friedel, T. Giamarchi, D. J\'erome,
J. P. Pouget, R. Shankar, J. Voit and many others.


\end{document}